\documentclass[dvipdfmx]{pasj01} 

\begin{document}
\SetRunningHead{Egusa et al.}{NIR and MIR images from {\it AKARI}/IRC pointed observations}

\title{
Revised calibration for 
near- and mid-infrared images from $\sim 4000$ pointed observations 
with {\it AKARI}/IRC
}

%

%
 \author{%
   Fumi \textsc{Egusa}\altaffilmark{1,2}
   Fumihiko \textsc{Usui}\altaffilmark{3}
   Kazumi \textsc{Murata}\altaffilmark{1}
   Takuji \textsc{Yamashita}\altaffilmark{1}
   Issei \textsc{Yamamura}\altaffilmark{1}
   and
   Takashi \textsc{Onaka}\altaffilmark{3}}
 \altaffiltext{1}{Institute of Space and Astronautical Science, 
Japan Aerospace Exploration Agency, 
Sagamihara, Kanagawa 252-5210, Japan}
 \altaffiltext{2}{National Astronomical Observatory of Japan, Mitaka, Tokyo 181-8588, Japan}
 \email{fumi.egusa@nao.ac.jp}
 \altaffiltext{3}{University of Tokyo, Bunkyo-ku, Tokyo 113-0033, Japan}

\KeyWords{methods: data analysis -- techniques: image processing -- infrared: general} 

\maketitle

\begin{abstract}
 The Japanese infrared astronomical satellite {\it AKARI} performed $\sim 4000$ pointed observations 
for 16 months until the end of 2007 August,
when the telescope and instruments were cooled by liquid Helium.
 Observation targets include solar system objects, Galactic objects, 
local galaxies, and galaxies at cosmological distances.
 We describe recent updates on calibration processes of 
near- and mid-infrared images taken by the Infrared Camera (IRC), 
which has nine photometric filters covering 2--27 $\mu$m continuously.
 Using the latest data reduction toolkit, 
we created calibrated and stacked images from each pointed observation.
About 90\% of the stacked images have a position accuracy better than $1.5''$.
 Uncertainties in aperture photometry estimated 
from a typical 
standard sky deviation
of stacked images are a factor of 
$\sim$ 2--4 smaller than those of AllWISE at similar wavelengths.
 The processed images together with documents such as process logs
as well as the latest toolkit are available online.
\end{abstract}

\section{Introduction}
 {\it AKARI} is the Japanese infrared (IR) satellite \citep{AKARI} 
launched on 
2006 February 21\footnote{
 All the dates in this paper are in UT.}
and operated until 
2011 November 24.
 The {\it AKARI} mission consists of several different phases of observations: 
a commissioning period prior to scientific operation called ``Performance Verification (PV)'' phase, 
followed by Phase 1 and 2 for scientific observations, the second PV phase, 
and Phase 3 for only near-IR (NIR) observations.
 Phase 1 started on 
2006 May 8
and lasted for half a year. 
 Most of time during Phase 1 was devoted to the all-sky survey 
at mid- and far-IR wavelengths.
 Phase 2 started on 
2006 November 10,
and many pointed observations 
together with supplemental all-sky survey 
were performed 
during this phase until 2007 August 26, 
as the cryogenic liquid Helium boiled off. 
 After the second PV phase to optimize the system performance under warmer environments, 
Phase 3 started on 
2008 June 1
\citep{Ona10b}.
 Pointed observations only 
at NIR wavelengths
were carried out until 
2010 February 15.
 No science observations were performed between then and the end of the {\it AKARI} operation.

\begin{figure}
 \begin{center}
  \includegraphics[width=\linewidth]{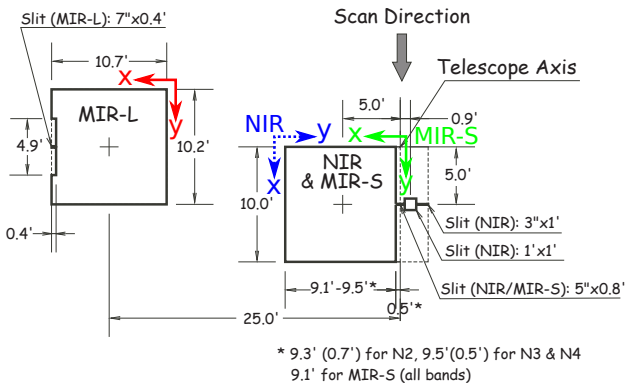} 
 \end{center}
\caption{The 
field of views
of the three channels of {\it AKARI}/IRC, revised from \citet{IRC}. 
The x- and y-axis directions of 
images from
each channel are indicated by arrows: 
blue dotted for NIR, green solid for MIR-S, and red solid for MIR-L.
The scan direction is along the ecliptic meridian.}
\label{fig:FoVs}
\end{figure}

 The Infrared Camera (IRC) is designed for observations 
at 
NIR and mid-IR (MIR)
wavelengths \citep{IRC}
while the Far-Infrared Surveyor (FIS) is for those at far-IR wavelengths \citep{FIS}.
 The IRC has three channels (NIR, MIR-S, and -L) and each channel 
has three photometric filters together with spectroscopic dispersers.
 These nine filters are named as N2, N3, N4, S7, S9W, S11, L15, L18W, and L24.
 The first letter indicates the channel, the numbers indicate
its representative wavelength in $\mu$m, and ``W'' denotes wide band.
 They
cover 2--27 $\mu$m continuously and are 
useful
to detect various objects at various redshifts 
and/or at various conditions.
 For example, emission and absorption features due to interstellar dust 
at NIR and MIR wavelengths can be used to trace 
interstellar medium at different conditions (e.g.\ \cite{Sak07}; \cite{Mori12}) 
and star-forming galaxies at different redshifts (e.g.\ \cite{Pea10}; \cite{Taka10}).
 The filter response curve, i.e.\ the wavelength coverage, of each filter 
is presented in \citet{IRC}.
 As illustrated in Figure \ref{fig:FoVs}, 
the field of view (FoV) of NIR and MIR-S almost coincides, 
while that of MIR-L is $\sim 20'$ away, 
each FoV being 
$\gtrsim 10'$ on one
side.

 {\it AKARI} pointed observations were carried out based on proposals 
classified as Large Survey, Mission Program, Open Time, 
and Director's Time.
 Their targets span 
a wide range of objects 
such as asteroids in the solar system (e.g.\ \cite{HaseS08,Mul14}), 
evolved stars in our Galaxy (e.g.\ \cite{Ita07,Ari11a}), 
interstellar dust in local and nearby galaxies (e.g.\ \cite{Kane08,Yama10,Egu13}), 
and distant galaxies (e.g.\ \cite{Taka10,Murata13}).
 Among them, the Large Magellanic Cloud 
(LMC)
and the North Ecliptic Pole (NEP) 
regions were extensively observed as the Large Survey projects 
\citep{Ita08,Kato12,Shimo13,NEPD,NEPW}.

 Each pointed observation is identified with  
one ObsID, which is a combination 
of targetID (7-digit number) and subID (3-digit number), e.g.\ 1234567\_123. 
 In most cases, subIDs are in the chronological order.
 There are a few exceptions due to observing conditions.
 One pointed observation ($\sim 10$ min excluding maneuvers) 
consists of nine or ten exposure cycles 
(as the term ``exposure frames'' used in \citet{IRC})
with filter changes and dithering between the cycles.
 The astronomical observation template (AOT) defines their combinations 
for different types of observations as summarized in Table \ref{tab:aot}.
 One exposure cycle consists of short and long exposures and 
the number and duration of these exposures depend on the channels and AOTs.
 Short exposure frames are useful for observing extremely bright sources 
and for identifying saturated pixels in long exposure frames.
 For more detail about the IRC and AOTs, see \citet{IRC}.
 Long exposure durations in Table \ref{tab:aot} are from \citet{Tana08}.

\begin{table*}
  \tbl{Summary of IRC AOTs for Phase 1\&2
  \label{tab:aot}}{
    \begin{tabular}{lcccccc}
      \hline
      Name & filters per channel & \multicolumn{2}{c}{long exposure frames per filter} & 
      \multicolumn{2}{c}{long exposure durations [sec]} & 
      dithering \\
      \cline{3-4} \cline{5-6}
      & & NIR & MIR-S and -L & NIR & MIR-S and -L & \\
      \hline
      IRC00 & 1 & 10 & 30 & 44.4 & 16.4 & no \\
      IRC02 & 2 & 4 & 12 & 44.4 & 16.4 & yes \\
      IRC03 & 3 & 3 & 9 & 44.4 & 16.4 & yes \\
      IRC04 & 1$+$spectrometer & 1 & 3 & 44.4 & 16.4 & no \\
      IRC05 & 1 & 5 & 30 & 65.5 & 16.4 & no \\
      \hline
    \end{tabular}}
\end{table*}
 
 Along with 
the scientific outcome mentioned above,
data processing techniques 
for IRC images
have been improved.
 Artifacts and point spread function (PSF) shapes for 
MIR were examined and characterized by \citet{Ari11}. 
 \citet{Tsu11} investigated and modeled a short-time variation of NIR dark current.
 \citet{Egu13} created a template for MIR dark current and for the earthshine light 
by combining images from neighbor observations.
 Temporal variations of MIR-S flat pattern were explored by \citet{Murata13}.

 In principle, a user needs to reduce the raw data 
using the toolkit provided by the {\it AKARI} team, but the data reduction 
often requires experience and knowledge in IR observations.
 In order to promote using {\it AKARI}/IRC data by a wide range of researchers, 
 we have recently included all of the abovementioned improvements 
into the toolkit  
and processed 
all the raw data sets from Phase 1\&2 pointed observations
except those of failed observations. 
 In addition to the raw data sets, 
processed and calibrated images along with documents and the latest toolkit have been 
released on 
2015 March 31.
 All of these products are available from 
the {\it AKARI} observers website\footnote{http://www.ir.isas.jaxa.jp/AKARI/Observation/}.
 With the advantage of the continuous wavelength coverage, 
these newly released {\it AKARI}/IRC data sets will enable us to explore new science cases
as well as the studies originally planned in the proposals.

 In this paper, we describe major revisions in the toolkit in \S \ref{sec:red}, 
present properties of processed and released data in \S \ref{sec:results}, 
and discuss remaining issues in \S \ref{sec:issues}.

\section{Data reduction}\label{sec:red}
\begin{figure}[!p]
 \begin{center}
 \includegraphics[trim=80 10 370 10,clip,angle=0,width=1.075\linewidth]{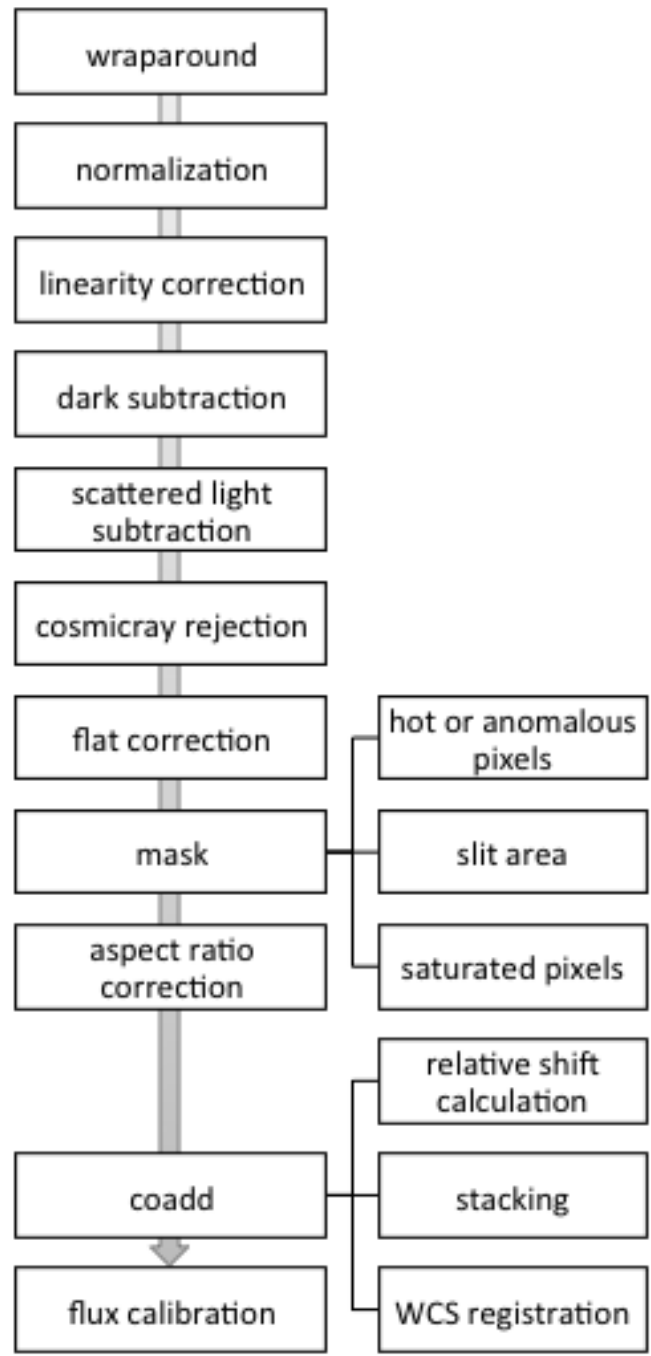} 
 \end{center}
\caption{
 Outline
of the standard pipeline processing flow 
in the IRC imaging toolkit.
Several preparative or minor steps are omitted for the sake of visibility.
}
\label{fig:toolkit}
\end{figure}

 The IRC imaging toolkit is based on 
the Image Reduction and Analysis Facility 
(IRAF\footnote{IRAF is distributed 
by the National Optical Astronomy Observatory, which is operated 
by the Association of Universities for Research in Astronomy, Inc., 
under cooperative agreement with the National Science Foundation.}) 
command language with additional tools written in C and Perl.
 Since the first release on 2007 January 4 (ver.\ 20070104, named after the release date), 
several major updates have been applied to the toolkit.
 The standard flow of the current pipeline processing (ver.\ 20150331) is 
outlined
in Figure \ref{fig:toolkit}.
 Several preparative or minor steps are not presented to simplify the flow.
 Among these processes, we here describe important steps that require 
a specific treatment due to the nature of {\it AKARI}/IRC images 
of pointed observations.

 Data presented in this paper are mostly products processed with the default setting.
 For example, 
we apply a sub-pixel sampling, so that one original pixel is divided into $2\times 2$ pixels.
 Signal in a pixel becomes $1/4$ of that of the original pixel after this sub-pixel sampling.
 We stack long-exposure frames only and 
use short-exposure frames just for identifying saturated pixels.
 We apply flux conversion factors from \citet{Tana08} to the stacked images.
 The unit of final pixel values is $\mu$Jy per pixel.
 Users can reprocess the raw data with different options, 
and skip and/or add some steps 
in order to obtain images best suited for their scientific aims.
 For more detail of the toolkit tasks and options, 
see the IRC data users manual  
available from the {\it AKARI} observers website.
 
\subsection{Dark frames}
 During a pointed observation, dark frames were obtained
before and after target observations, called pre- and post-dark frames, respectively.
 For Phase 1\&2, we identify three types of temporal variation in IRC dark frames: 
(i) short-term (i.e.\ within one pointed observation or $\sim 10$ minutes), 
(ii) intermediate-term (i.e.\ over several pointed observations or a few hours), and 
(iii) long-term (i.e.\ over the entire observation period or more than a few months).
 For the short-term variation, we measure the average level of 
the masked area for spectroscopic observations 
on each frame during an observation.
 Since pixels in this area are masked during imaging observations, 
the average level of these pixels should correspond to the average dark current level.
 The average level of the same area of the dark frame is also measured 
and their difference between the object and dark frames (i.e.\ a constant)
is subtracted or added accordingly 
when the dark frame is subtracted from an object frame.
 In the following, the latter two variations are described 
together with hot pixels in MIR frames.
 
 The default setting of the toolkit is to use a model by \citet{Tsu11} for NIR 
and neighbor dark frames for MIR long exposure frames.
 We use these default dark frames to calibrate images presented in this paper, 
i.e.\ the released data sets.

\subsubsection{Model for NIR dark frames} 
 The intermediate-term variation of dark current 
has been thought to be due to a passage of South Atlantic Anomaly.
 \citet{Tsu11} investigated this variation for NIR long-exposure frames 
and found that the variation cannot be fully calibrated 
by just adding or subtracting the constant. 
 Using more than 4000 pre-dark frames taken in Phase 1\&2, 
they created a model of this dark current variation for each pixel of the NIR array.
 An accuracy of the model was estimated to be $\sim 1$ ADU, 
which is about 10\% of the typical dark current 
and corresponds to $\sim 0.3~\mu$Jy for NIR long-exposure frames
\citep{Tana08}.

\subsubsection{Neighbor dark frames for MIR}
 The long-term variation appears as an increase of hot pixels 
and is more evident at longer wavelengths.
 Combining pre-dark frames in each pointed observation (called self-dark) 
better calibrates this variation than super-dark, 
which was created by combining $\sim 100$ pre-dark frames 
taken in the early phase of the satellite operation.
 However, during a standard pointed observation, only three pre-dark 
frames were obtained for MIR long exposure, 
so that the signal-to-noise (S/N)
ratio of self-dark is lower than that of super-dark.

\begin{figure}
 \begin{center}
 Observation date = 2006 May 31\\
  \includegraphics[trim=0 0 10 0,clip,width=0.475\linewidth]{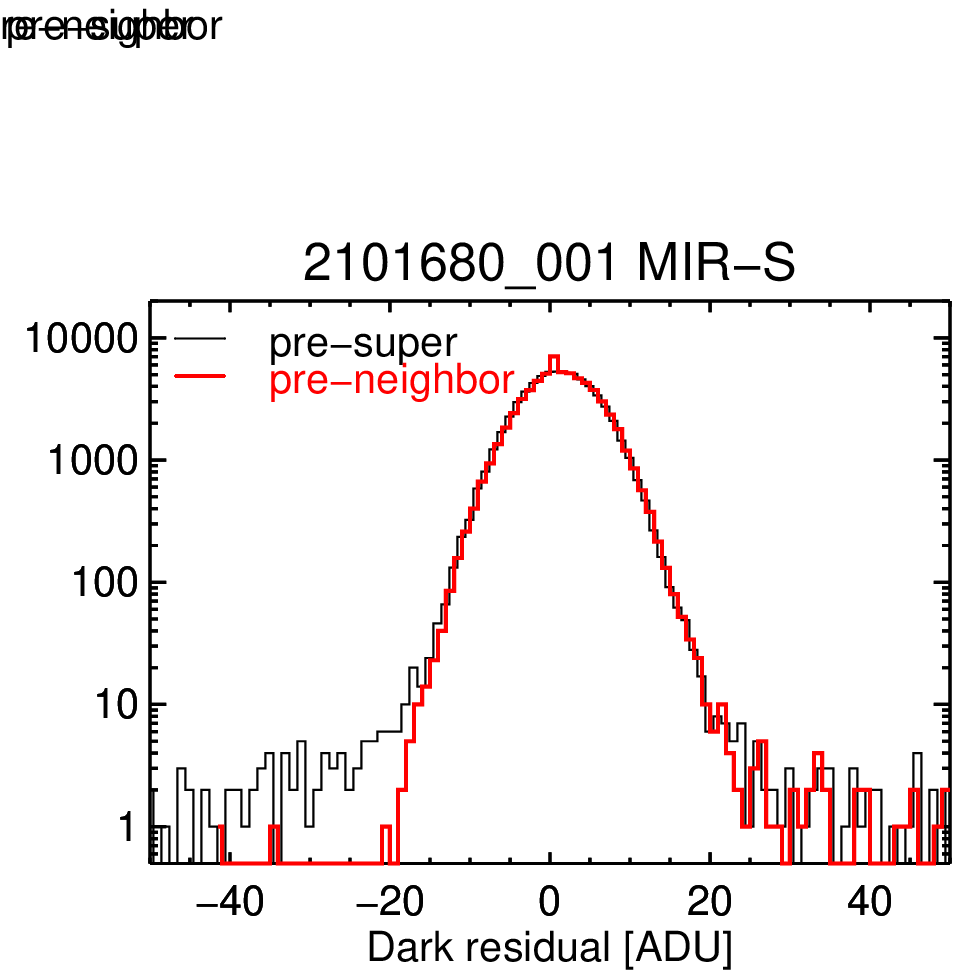} 
  \includegraphics[trim=0 0 10 0,clip,width=0.475\linewidth]{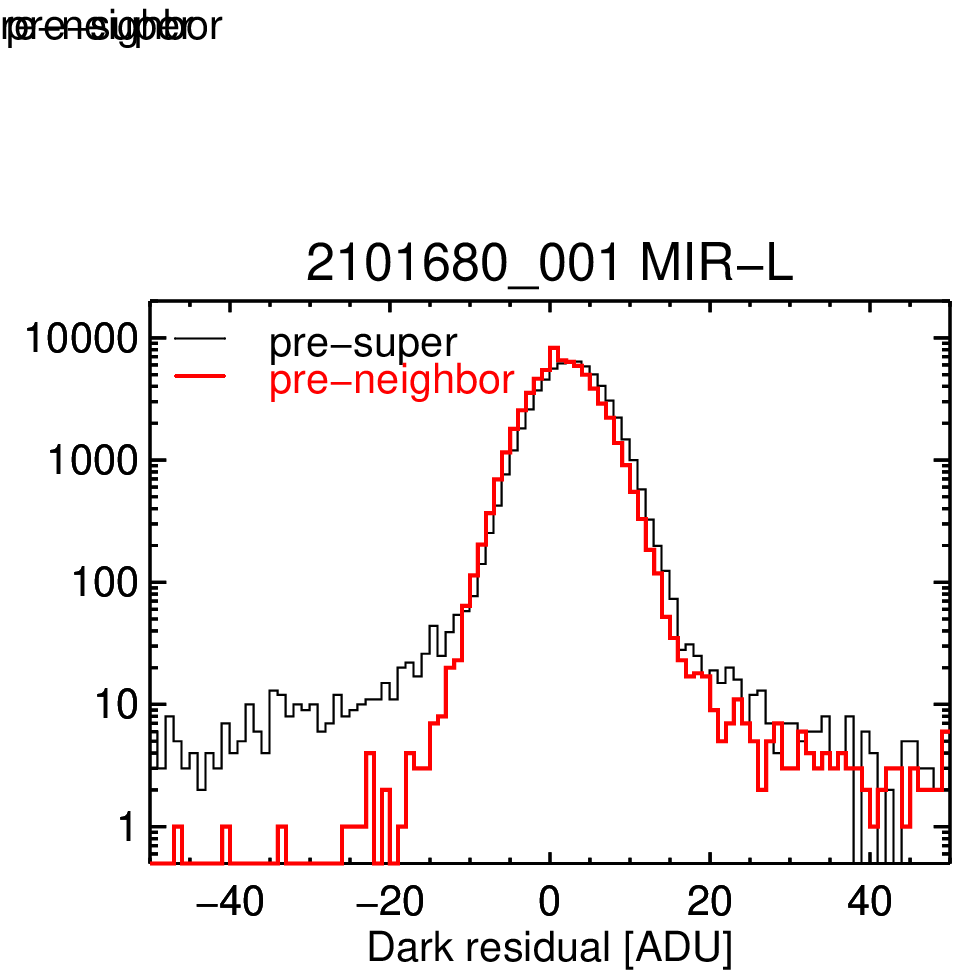} 
  Observation date = 2007 July 16\\
  \includegraphics[trim=0 0 10 0,clip,width=0.475\linewidth]{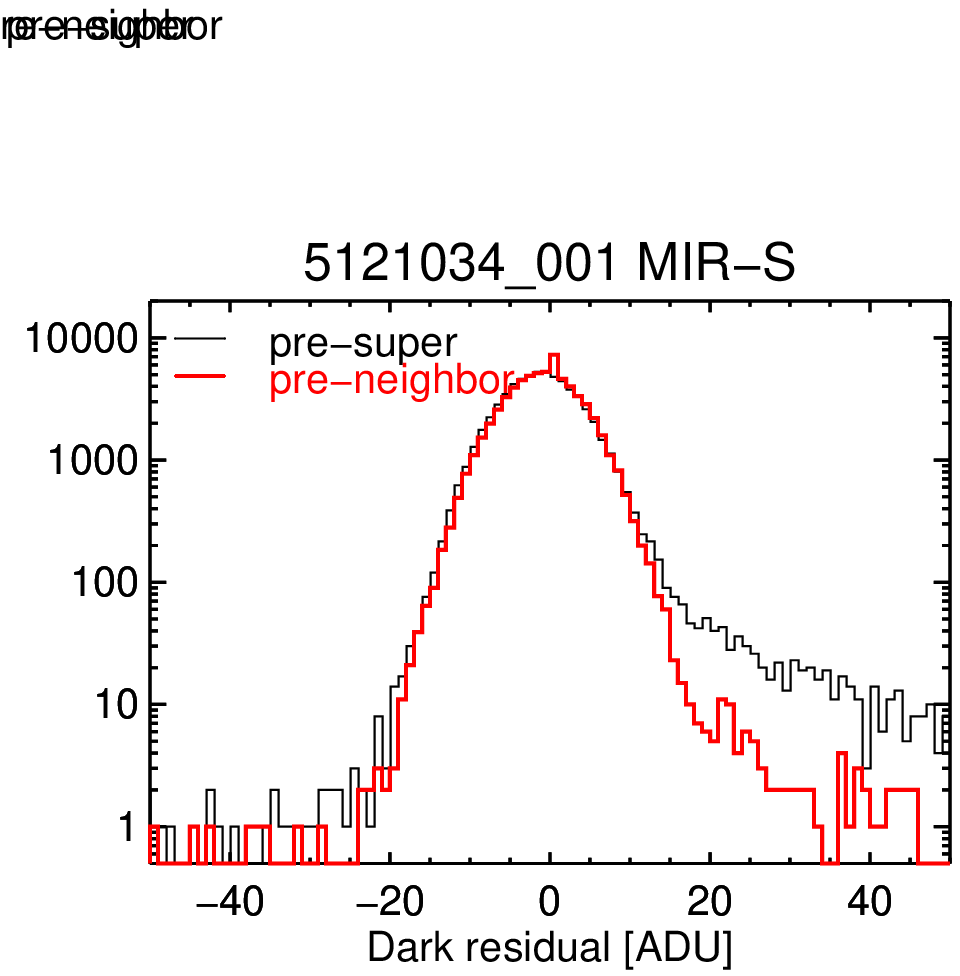} 
  \includegraphics[trim=0 0 10 0,clip,width=0.475\linewidth]{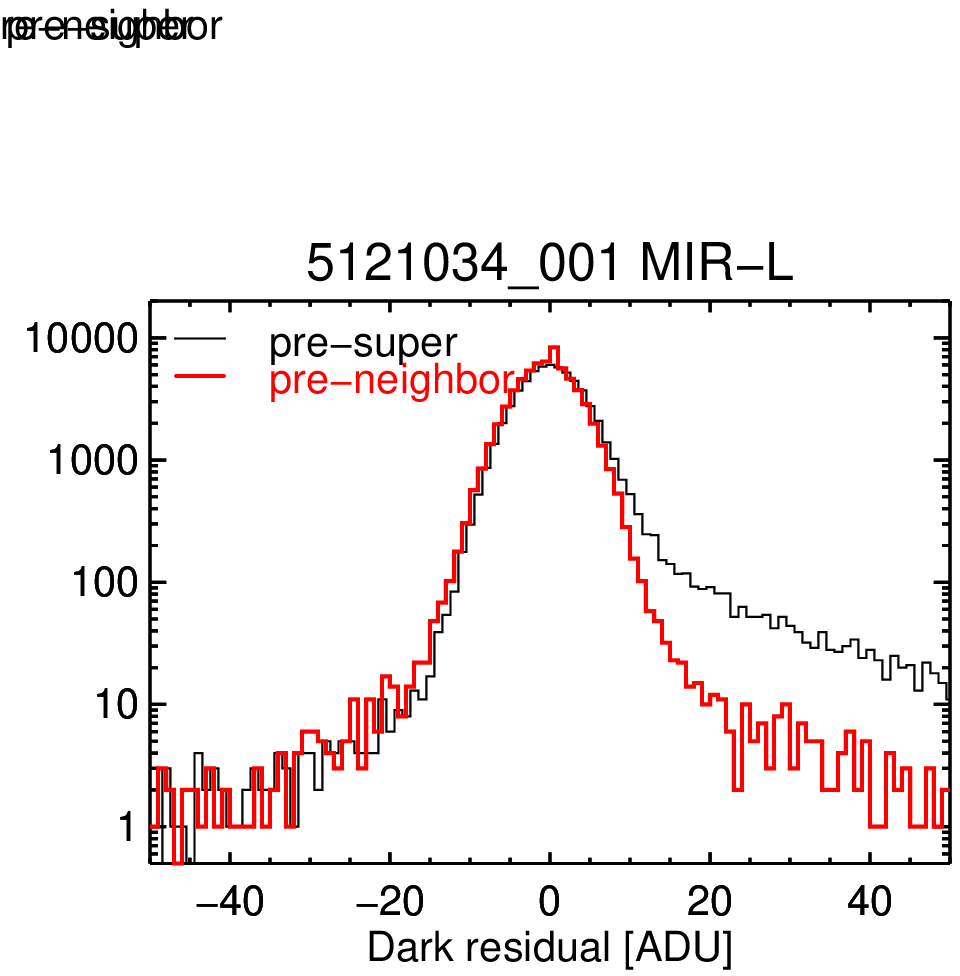} 
 \end{center}
\caption{
Dark residual [ADU] measured as pre-dark $-$ super-dark (thin black) 
or pre-dark $-$ neighbor-dark (thick red) frames. 
Left and right columns are for MIR-S and -L, respectively.
Top and bottom rows are for observations performed 
around the beginning and the end of Phase 1\&2, respectively.
ObsIDs and channels are presented in the top of each plot.
}
\label{fig:comp_dark}
\end{figure}

 In order to obtain a dark frame which represents the long-term variation 
with a high S/N, \citet{Egu13} and \citet{Murata13} combine pre-dark frames 
of neighbor pointed observations.
 Following 
this
strategy, for each ObsID, we combine pre-dark frames from 
five observations before and after that observation (i.e.\ 11 observations in total)
to create a neighbor dark frame of MIR-S and -L long exposure.
 The number of frames combined is 33 or more and a typical duration is about one day.
 By subtracting super or neighbor dark from pre-dark frames, 
we estimate uncertainties in the dark subtraction process.
 In Figure \ref{fig:comp_dark}, histograms of these MIR dark residuals
are presented for observations performed 
on 2006 May 31 (top) and 2007 July 16 (bottom), 
i.e.\ around the beginning and the end of Phase 1\&2.
 While the main part of gaussian profiles peaking around zero 
does not change with time, residuals of the super-dark subtraction (thin black line) 
have a significant positive tail compared to the neighbor dark subtraction (thick red line)
at the later stage of Phase 2 (bottom panels).
 This result indicates that many hot pixels are not fully corrected after subtracting the super dark.
 In the top panels, on the other hand, a tail of the black histogram appears on the negative side.
 It is most likely due to the fact that frames used to create the super dark 
include those taken after the observation shown in the top panels.
 Some pixels in the super dark may be affected by hot pixels appearing after 
this observation.
 As a result, such pixels are over-subtracted after the super-dark subtraction.
 Nevertheless, it is clear from this figure
that neighbor dark frames are more appropriate 
than super dark frames for reducing the effect of hot pixels.
 We estimate the uncertainty of dark subtraction is $\sim 5$ and 4 ADU for 
MIR-S and -L, respectively, from the width of the main gaussian profiles.
 These uncertainties are about 10\% of the typical dark current 
and correspond to $\sim 4$ and $10~\mu$Jy for MIR-S and -L long-exposure frames,
respectively \citep{Tana08}.

\subsubsection{MIR hot pixels from neighbor dark frames}
\begin{figure}
 \begin{center}
  \includegraphics[width=\linewidth]{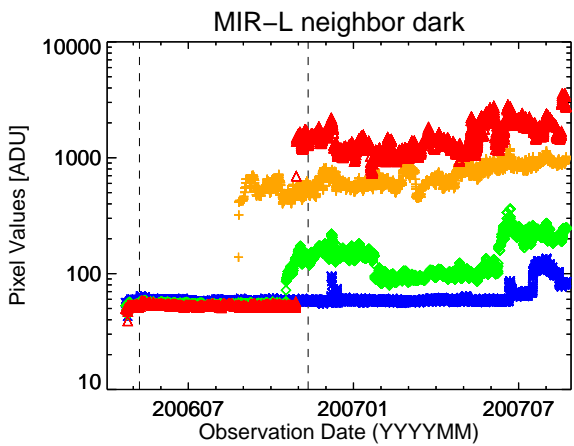} 
 \end{center}
\caption{
Pixel values [ADU] of four selected pixels in MIR-L neighbor dark frames
indicated by different colors.
The horizontal axis is observation dates in YYYYMM format.
Two vertical dashed lines indicate the start of Phase 1 and 2.
}
\label{fig:trace_dark}
\end{figure}

\begin{figure}
 \begin{center}
  \includegraphics[width=\linewidth]{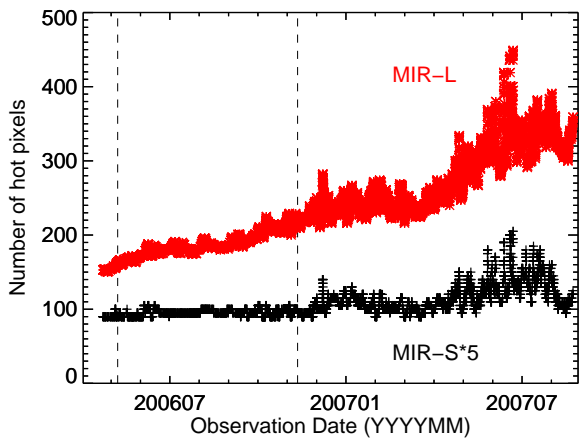} 
 \end{center}
\caption{The number of hot pixels identified in neighbor dark frames for 
MIR-S (black plus, multiplied by 5) and -L (red asterisk).
The horizontal axis and the vertical lines are the same as Figure \ref{fig:trace_dark}.
}
\label{fig:nhot}
\end{figure}

 From these neighbor dark frames, hot pixels are defined as pixels whose values  
exceed a certain threshold.
 In the toolkit, the default threshold is $500.0$ ADU and users can adjust it if necessary.
 In order to determine this default value, we investigated a temporal variation 
of representative pixels in MIR-L neighbor dark frames, as shown in Figure \ref{fig:trace_dark}.
 We found that once a pixel value exceeds $\sim 500$ ADU (i.e.\ orange and red points in the figure)
it stays at the same high level in most cases 
and its fluctuation becomes larger than the typical dark current, which is $\sim 50$ ADU for MIR.
 On the other hand, when a pixel value is $\lesssim 100$--$200$ (i.e.\ green points in the figure), 
its fluctuation is smaller or comparable to the typical dark current 
(Note that y-axis of this figure is logarithmic scale).
 We thus regard such a pixel can be calibrated by the dark subtraction
and set the default threshold to be $500.0$ ADU.
 The number of hot pixels is plotted against observation dates in Figure \ref{fig:nhot}.
 
\subsection{Flat pattern}
 The flat pattern is a distribution of pixel sensitivity on a detector.
 We create a flat frame for each filter by combining object frames (i.e.\ sky flat).
 The default setting of the toolkit is to subtract a constant sky
after dividing by a flat frame.
 In the following part of this subsection, we explain special treatments 
needed in the flat correction.

\subsubsection{Extended ghosts in MIR-L}\label{sec:MIRflat}
 \citet{Ari11} investigated ghost patterns in MIR-S and -L arrays 
and successfully separated extended ghost (or artificial flat) patterns 
and ``true'' flat patterns for 
L15 and L24.
 Using these two ``true'' flat patterns
for L15 and L24, \citet{Murata13} created a ``true'' flat pattern
for L18W.
 In the toolkit, 
a difference between observed and ``true'' flat patterns of each MIR-L filter
is treated as an extended ghost pattern 
and is subtracted from object frames during the flat correction 
instead of subtracting a constant sky.

\subsubsection{The ``soramame'' pattern in NIR and MIR-S}\label{sec:soramame}
 It has been known that in the bottom right corner of the MIR-S FoV 
a noticeable pattern was present until 
2007 January 7.
 From its shape, this pattern is called ``soramame'' (broad bean in Japanese).
 As a faint but similar pattern was also seen in the bottom left corner of the NIR FoV, 
its cause is thought to be an obstacle in the light path before 
the beam splitter.
 The effect of ``soramame'' pattern is typically 
a few percent for NIR 
and up to 10 percent for MIR-S.
 In addition, the sky background is less bright in NIR, 
so that ``soramame''-shaped artifact is more prominent in MIR-S.

 \citet{Murata13} investigated a temporal variation of the MIR-S ``soramame'' shape in detail 
and found five periods with different shapes.
 Since the shape was not stable even within one period, 
\citet{Murata13} created a ``soramame'' pattern for each frame 
from neighbor frames. 
 However, the number of neighbor frames is not always enough to 
create a neighbor flat for the entire period of Phase 1\&2.
 We thus re-defined the five periods into p1, p23, p4, and p5, and created a flat frame 
for each MIR-S filter and each period.
 The period p23 is a combination of the second and third periods 
defined by \citet{Murata13}.
 We combined these two periods, since the variation during these periods was small and gradual 
and thus it was hard to 
draw a dividing line.
 It is consistent with the fact that typical patterns for these two periods 
presented by \citet{Murata13} (``ii'' and ``iii'' in their Figure 3 (a)) resemble each other.
 Furthermore, the earthshine light effect (described in \S \ref{sec:EL}) is 
significant during the first half of p23.
 Since we create a flat frame for each period by stacking object frames taken in that period, 
frames with artificial patterns fixed to the detector coordinates (such as the earthshine light) 
should not be used.
 We thus created a flat frame from data only in the latter half of p23 
and the toolkit applies it to all the data in p23.
 The starting date of each period is listed in Table \ref{tab:periods} 
and the last period, p6, corresponds to the period without ``soramame''.

 For NIR, the number of frames was smaller and the temporal variation in shape of 
the artifact was less clear.
 We thus created one flat frame for each filter for all the periods with ``soramame'' 
instead of splitting into four periods.

\begin{figure}
 \begin{center}
  \includegraphics[trim=0 60 0 0,clip,width=\linewidth]{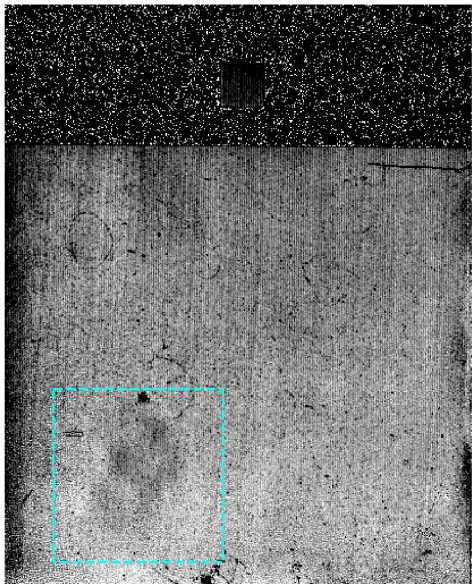} 
 \end{center}
\caption{A flat frame for N4
to demonstrate ``soramame''-shaped artifact within
the cyan dashed box, which encloses the same area 
as those in Figure \ref{fig:S7flat} 
(Note that the bottom left corner in NIR corresponds to 
the bottom right corner in MIR-S.
See Figure \ref{fig:FoVs} for the FoV alignment and axis directions.).
}
\label{fig:N4flat}
\end{figure}

\begin{figure*}
 \begin{center}
  \includegraphics[trim=0 50 0 0,clip,width=\linewidth]{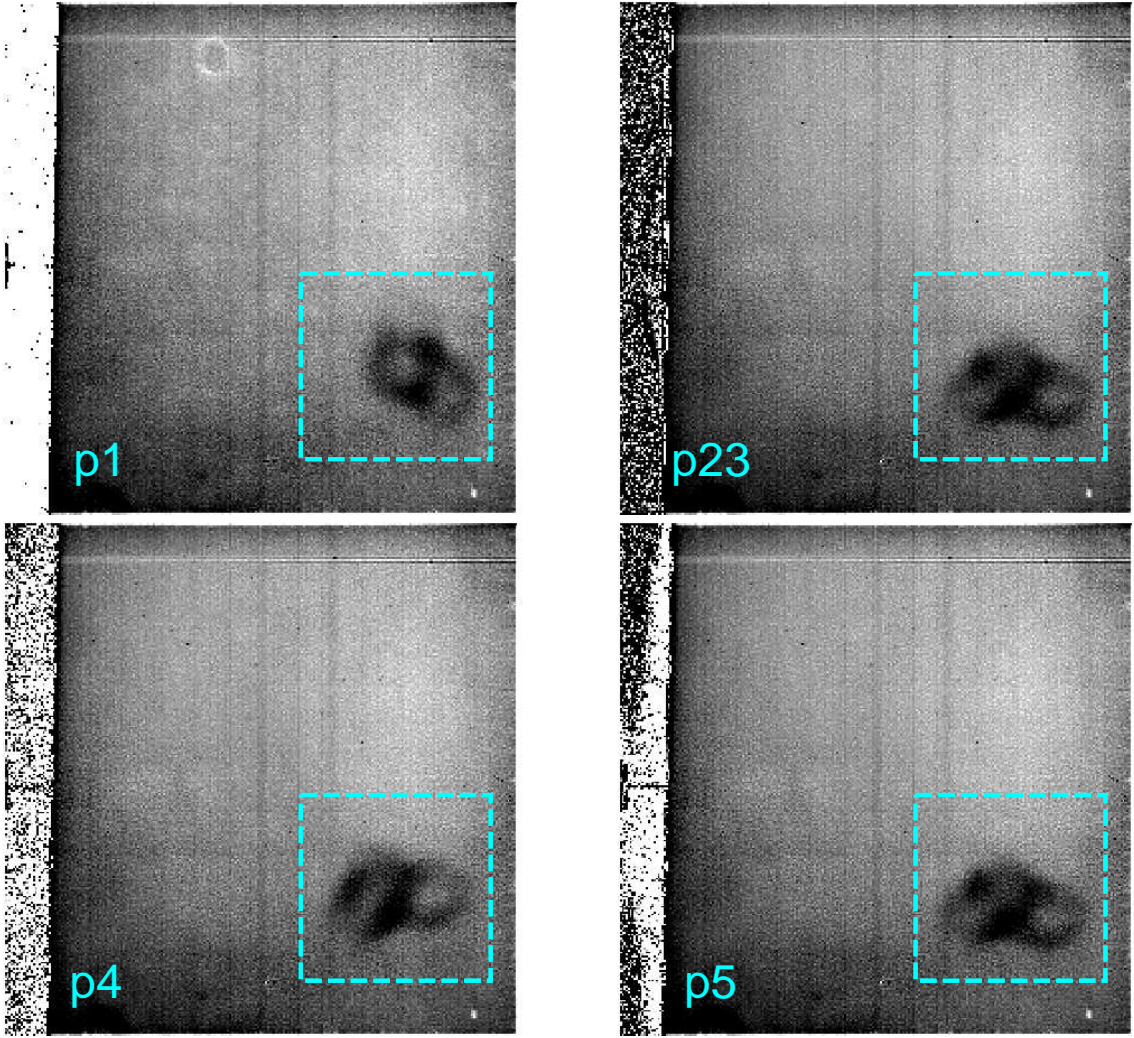} 
 \end{center}
\caption{Flat frames for S7 with ``soramame'' 
within the cyan dashed boxes in the bottom right corner,
which enclose the same area 
as that in Figure \ref{fig:N4flat} 
(Note that the bottom left corner in NIR corresponds to 
the bottom right corner in MIR-S.
See Figure \ref{fig:FoVs} for the FoV alignment and axis directions.).
From top left to bottom right, the periods are p1, p23, p4, and p5 
as indicated in the bottom left corner of each panel.
 The white circle in the top left of the p1 flat frame is a residual of an observed source 
due to the small number of frames available during this period.
 See \S \ref{sec:soramame} for more detail.
}
\label{fig:S7flat}
\end{figure*}

 Flat frames with ``soramame'' are presented in Figures \ref{fig:N4flat} 
and \ref{fig:S7flat} for N4 and S7, respectively.
 The cyan dashed boxes in these two figures 
enclose the same area on the focal plane 
(see Figure \ref{fig:FoVs} for the FoV alignment of NIR and MIR-S).
 A bright circle in the p1 S7 flat is a residual of an observed source.
 Since the p1 period was short, the number of frames is not enough 
to remove such residuals and thus the reliability and S/N of the flat for this period are low.
 We thus decided not to include flat frames for p1 in the toolkit
and thus not to deliver the processed data from this period.
 Note that p1 is in the PV phase.

\begin{table}
  \tbl{Starting date of each period for ``soramame'' artifact
  \label{tab:periods}}{
    \begin{tabular}{lc}
      \hline
      Name & Date (YYYY-MM-DD) \\
      \hline
      p1 & 2006-04-22 \\
      p23 & 2006-04-29 \\
      p4 & 2006-12-09 \\
      p5 & 2006-12-15 \\
      p6 & 2007-01-07 \\
      \hline
    \end{tabular}}
\end{table}


\subsection{Relative shift between frames}\label{sec:calcshift}
 The position of FoV on the sky is not always the same during one pointed observation 
due to an intentional dithering 
and/or unintentional jittering and drifting of the satellite's attitude.
 For each filter, the shift values in x- and y- direction and the 
counter-clockwise
rotation angle 
relative to the first frame of a pointed observation are calculated using 
bright sources within the FoV.
 The minimum number of the sources used in this calculation is set to seven.  
 In other words, if only six or less sources are found in a frame, 
no calculation is performed and this frame is excluded from stacking.

\begin{figure*}
 \begin{center}
\includegraphics[width=0.3\linewidth]{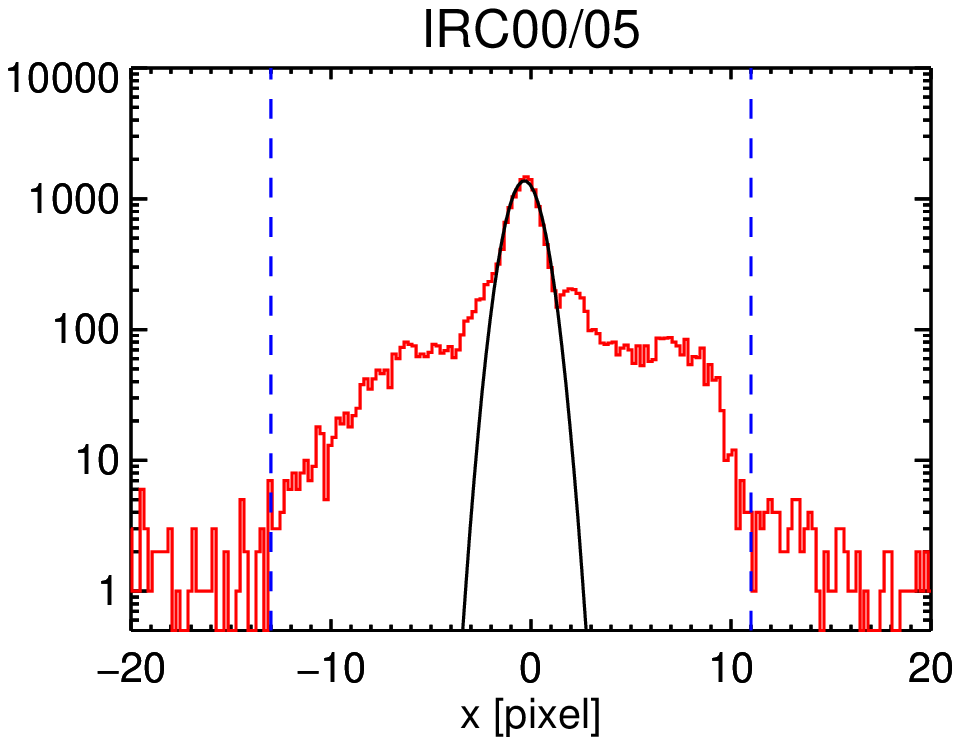} 
\includegraphics[width=0.3\linewidth]{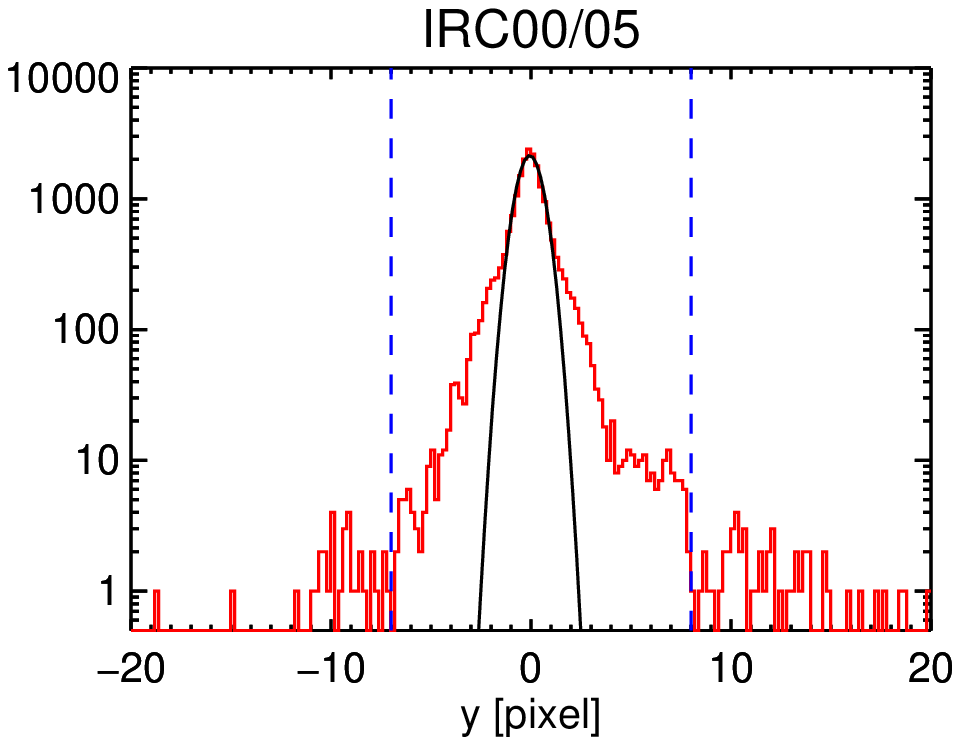} 
\includegraphics[width=0.3\linewidth]{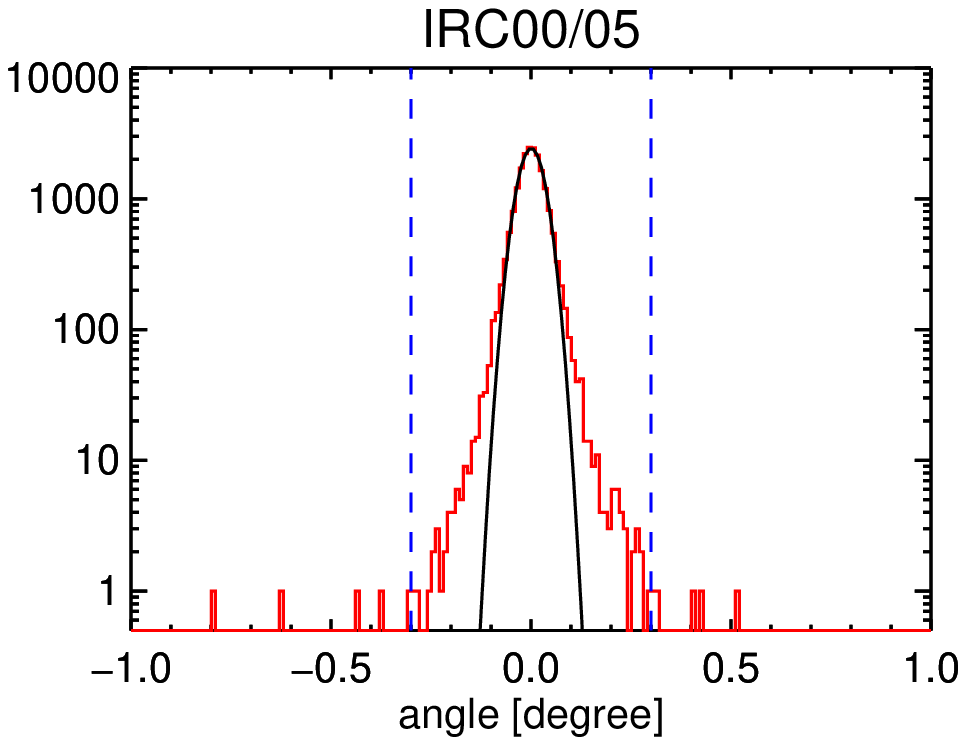} 
 \end{center}
\caption{
Histogram of relative shift and rotation angle of MIR-S frames for IRC00 and IRC05, 
i.e.\ observations without dithering.
Pixel 
numbers
are after the sub-pixel sampling, i.e.\ twice the original value.
One pixel is $1.17''$ for MIR-S.
Black curves represent the results of gaussian fit to the histogram 
and blue dashed lines indicate the lower and upper limits adopted in the toolkit.}
\label{fig:hshift}
\end{figure*}

 For observations without dithering (AOT=IRC00 or IRC05),
histograms of these shifts and angle calculated for MIR-S frames 
are presented in Figure \ref{fig:hshift} to demonstrate the pointing stability.
 Histograms for those with dithering (AOT=IRC02 or IRC03) 
are presented in Appendix \ref{sec:stat_0203}. 
 Note that the shift values are calculated after the sub-pixel sampling, 
so that two pixels in the plot correspond to one original pixel.
 We fit a gaussian profile to each histogram and find that 
the histogram for the shift in x-direction has significant wings. 
 This result indicates that the drift was mostly along the x-axis direction 
of MIR-S images 
(see Figure \ref{fig:FoVs} for the axis direction),
which is consistent with the measured PSF sizes 
described in 
\S \ref{sec:PSF}. 
 The drift issue is also discussed in 
\S \ref{sec:drift}.

 Upper and lower limits to the shift values are set to exclude false matching 
for MIR-S and -L frames. 
 We manually determine these limits to include most of the wings of the histograms
and to be at where the histogram profiles drop steeply.
 These limits are indicated by blue dashed lines in 
Figure \ref{fig:hshift},
and given in a constant parameter file of the toolkit.
 Meanwhile, no such limit is employed for NIR frames.
 We notice that some of the frames are excluded due to their large drifts 
rather than the false matching.
 However, the current toolkit cannot distinguish these two factors, since it 
only checks shift values of one frame against the corresponding limits. 
 In some cases, the telescope was drifting gradually, 
so that relative shift values {\it between adjacent frames} are small
although those {\it with respect to the first frame} exceed the limit.
 These frames can be identified and resurrected by checking 
difference in shift values between adjacent frames, 
but such a scheme is not established yet.
 
\subsection{WCS matching}
 The World Coordinate System (i.e.\ Right Ascension and Declination) 
information from the satellite telemetry was not accurate enough 
for scientific purposes because of the insufficient absolute accuracy of 
the attitude and orbit control system 
and the pointing stability.
 The toolkit 
determines
the WCS 
of a stacked image by matching detected sources in the image 
and sources in the 2MASS \citep{2MASS} catalog for NIR 
or in the Wide-field Infrared Survey Explorer (WISE; \cite{WISE}) catalog for MIR-S and -L.
 We note that the toolkit performs this WCS matching process 
for each stacked image (i.e.\ each filter) due to the following reasons: 
(i) more sources are available in stacked images compared to individual frames 
(i.e.\ WCS matching to individual frames is more difficult), and
(ii) the FoVs of NIR/MIR-S and MIR-L are $\sim 20'$ apart
and their absolute roll angle is not confirmed to be stable enough 
(i.e.\ WCS information from MIR-S cannot be simply transferred to MIR-L 
or absolute positions for all the sources found in all the images for one pointed observation 
cannot be solved all at once). 
 The matching tolerance is set to be $1.5''$ at all the wavelengths.
 
 Typical uncertainties in this process are $0.4''$, $0.5''$, and $0.8''$, 
for NIR, MIR-S, and -L, respectively 
(See \S \ref{sec:results_wcs} and Appendix \ref{sec:wcserr} for more detail).

\section{Results from all-data processing}\label{sec:results}
 In this section, we describe how the released data sets were created 
and their quality such as a position accuracy, image sensitivity, and PSF.

\subsection{Process summary}\label{sec:procsum}
\begin{figure*}
 \begin{center}
  \includegraphics[trim=0 25 0 0,clip,width=\linewidth]{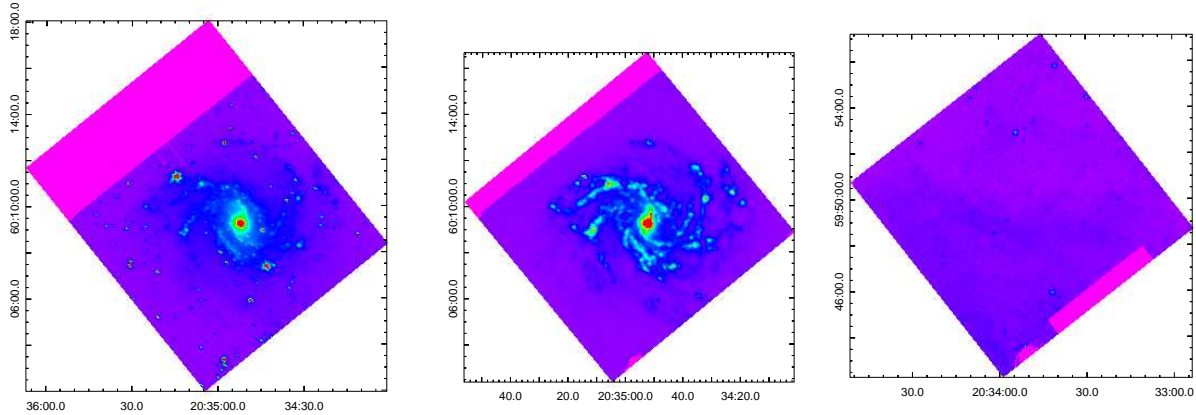} 
 \end{center}
\caption{Sample of calibrated and stacked images (N3, S11, and L15 from left to right) 
from a pointed observation (ObsID=1400620\_001)
toward a nearby spiral galaxy NGC 6946 in the NIR and MIR-S FoVs. 
The FoV of MIR-L image
is $\sim 20'$ away.
Coordinates are R.A.\ and Dec.\ (J2000).
The flux ranges presented are [-10,100], [-100,1000], and [-20,200] $\mu$Jy/pix 
for N3, S11, and L15, respectively.
Magenta area are masked pixels mostly due to the slit for spectroscopic observations.}
\label{fig:sample}
\end{figure*}

 As already mentioned, 
the sub-pixel sampling is performed and  
only long exposure frames are stacked.
 The unit of pixel values is $\mu$Jy per pixel, with the pixel size of 
$0.723''$, $1.17''$, and $1.19''$, for NIR, MIR-S, and -L, respectively.
 The Right Ascension and Declination (J2000) of stacked images are available 
when the WCS matching is successful.
 The background sky is subtracted before stacking.
 All the options and parameter settings adopted to create the released data sets 
are recorded in a log file included in a package for each ObsID (See also Appendix \ref{sec:datasets}).
 A sample of processed images for a pointed observation is 
presented in Figure \ref{fig:sample}.
 
 During 
the first PV and
 Phase 1\&2, $\sim 4000$ pointed observations were performed with the IRC.
 When
the FIS was the primary instrument to observe a target, 
the IRC was operated 
and observing at a field $\sim 20'$ away from that of the FIS.
 These observations are called parallel observations.
 The FIS was operated either in the slow-scan mode for imaging observations 
(AOT = FIS01 or FIS02) 
or in the staring mode for spectroscopic observations (AOT = FIS03).
 The parallel IRC observations of the latter (i.e.\ FIS03) are denoted as 
IRC05, and are included in the released data set.
 One exposure cycle with photometric filters taken as positional references 
in the spectroscopic observations (IRC04) is also included.
 On the other hand, p1 is excluded due to the low S/N of MIR-S flat frames 
as explained in \S \ref{sec:soramame}.
 We also exclude several observations without any valid data.

\begin{figure}
 \begin{center}
  \includegraphics[width=\linewidth]{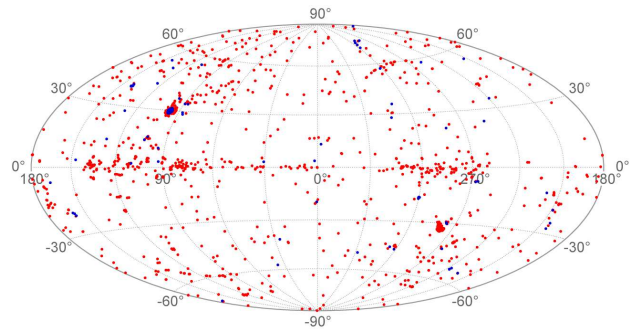} 
 \end{center}
\caption{All the pointed observations during Phase 1\&2 in the galactic coordinates.
Red and blue dots indicate imaging and spectroscopic observations, respectively.}
\label{fig:allID}
\end{figure}

 The number of observations (i.e.\ ObsIDs) for the standard observation modes 
(with each AOT and AOTparameter) is listed in Table \ref{tab:nIDs}.
 The AOTparameter, N or L, denotes whether the main target is in the 
NIR and MIR-S FoVs or in the MIR-L FoV.
 Figure \ref{fig:allID} illustrates the on-sky distribution of imaging (red) 
and spectroscopic (blue) observations during Phase 1\&2.
 It is clearly seen that many observations are made 
toward the Galactic plane ($b\sim 0^\circ$), 
NEP ($l\sim 100^\circ$, $b\sim 30^\circ$), and LMC ($l\sim 280^\circ$, $b\sim -30^\circ$).
 The number of observations for each proposal is presented in Appendix \ref{sec:datasets}.
 
\begin{table}
\tabcolsep16pt 
  \tbl{The number of observations for the standard observation modes in Phase 1\&2
  \label{tab:nIDs}}{
    \begin{tabular}{lcc}
\hline\noalign{\vskip3pt} 
     AOT$^{*}$ & N$^{\dag}$ & L$^{\dag}$  \\    [2pt] 
\hline\noalign{\vskip3pt} 
      IRC00 & 12 & 0 \\
      IRC02 & 1189 & 370 \\
      IRC03 & 542 & 151 \\
      IRC04 & 650 & 230 \\
      IRC05 & 513 & 252 \\[2pt] 
\hline\noalign{\vskip3pt} 
    \end{tabular}}
 \begin{tabnote}
\hangindent6pt\noindent
\hbox to6pt{\footnotemark[$*$]\hss}\unskip%
 Observational settings for each AOT are summarized in Table \ref{tab:aot}.\\
\hangindent6pt\noindent
\hbox to6pt{\footnotemark[$\dag$]\hss}\unskip%
 N and L are the AOT parameter specifying a target in NIR/MIR-S and -L FoVs, respectively.
 See \S \ref{sec:procsum} for detail.
 \end{tabnote}
\end{table}

 Since the exposure cycle is the same for MIR-S and -L, 
the relative shift values for MIR-S and -L should also be the same.
 Meanwhile, the shift calculation for MIR-L is generally more difficult 
as the number of bright sources at longer wavelengths is smaller.
 For imaging 
data, 
we also created stacked images by using
the ``coaddLusingS'' scheme, 
in which shift values calculated for MIR-S frames are used to stack MIR-L frames.
 Among these two stacked images for one MIR-L filter, we selected the better one to be released
based on the number of frames used for stacking and the results of WCS matching.
 In the released data sets, 
the fraction of stacked images by the ``coaddLusingS'' scheme is 
56\%, 70\%, and 78\% for L15, L18W, and L24, respectively.
 These fractions are consistent with the general picture mentioned above
(i.e.\ more sources brighter at shorter wavelengths) but not 100\%.
 We deduce that this is due to the different FoVs for MIR-S and -L 
-- when a target is in the MIR-L FoV, the number of bright sources 
can be larger for MIR-L, resulting in a better result with the original MIR-L shift values.

\subsection{Position accuracy}\label{sec:results_wcs}
 Stacked images with successful WCS matching are defined as ``Matched Images''.
 Typical uncertainties in the WCS matching are $0.4''$, $0.5''$, and $0.8''$
for NIR, MIR-S, and MIR-L, respectively.
 See Appendix \ref{sec:wcserr} 
for more detail.
 We examined all of the Matched Images visually 
and rejected an image if most of the sources in that image were displaced 
$>5''$ from catalogued sources.
 We did not take into account elongated PSFs 
(see \S \ref{sec:PSF} and also \S \ref{sec:drift}) for rejection.
 Rejected images amount to $\sim 1$\% of the Matched Images, and
are re-classified as ``Failed Images'' together with unsuccessful WCS matching.
 The numbers of Matched and Failed Images for each filter 
are summarized in Table \ref{tab:wcsrate}.
 The success rates are calculated as 
Matched/(Matched+Failed)
and $\sim 98$\% for NIR and $> 90$\% for MIR-S.
 The rate generally decreases with wavelength 
and is $\sim 50$\% for L24.
 The average success rate for all the stacked images is 87\%.

\begin{table*}
  \tbl{Results of WCS matching
  \label{tab:wcsrate}}{
    \begin{tabular}{lcccccccccc}
      \hline
       & N2 & N3 & N4 & S7 & S9W & S11 & L15 & L18W & L24 & All\\
      \hline
Matched Images  & 850 & 2545& 1772& 2407& 1694& 2293& 2331& 1327& 1314 & 16533\\
Failed Images   & 14  & 58  & 34  & 258 & 70  & 193 & 339 & 404 & 1185 & 2555\\
Success Rate & 0.98& 0.98& 0.98& 0.90& 0.96& 0.92& 0.87& 0.77& 0.53 & 0.87\\
      \hline
    \end{tabular}}
\end{table*}

\subsection{Sensitivity}\label{sec:sen}
 A standard deviation of a background sky signal in
stacked images is estimated using data sets 
from the NEP observations 
performed during 
2006 August and 2007 April,
when the earthshine effect (discussed in 
\S \ref{sec:EL}) was not significant.
 A temporal variation for the whole Phase 1\&2 
is presented in Appendix \ref{sec:sigma_var}.
 For each stacked image, we create a histogram of pixel values and fit a gaussian profile.
 A typical value of 
the gaussian profile width,
$\sigma$, is adopted as the 
standard sky deviation.
 Table \ref{tab:skyrms} lists this value for each filter and AOT
in units of $\mu$Jy/pixel.
 Values for IRC02 are not listed as this AOT was not used for the NEP observations, 
but they should be in between those of IRC03 and IRC05.
\begin{table*}
  \tbl{Typical standard sky deviation of stacked images from NEP observations [$\mu$Jy/pix]
  \label{tab:skyrms}}{
    \begin{tabular}{lccccccccc}
      \hline
      AOT & N2 & N3 & N4 & S7 & S9W & S11 & L15 & L18W & L24 \\
      \hline
      IRC03 & 0.11 & 0.079 & 0.078 & 0.65 & 0.70 & 1.1 & 2.0 & 1.9 & 4.0 \\
      IRC05 & 0.061 & 0.043 & 0.042 & 0.38 & 0.54 & 0.75 & 1.2 & 1.1 & 2.6 \\
      \hline
    \end{tabular}}
 \begin{tabnote}
 Measured in stacked images created with sub-pixel sampling.
 \end{tabnote}
\end{table*}

 Note that these values are for the stacked images, i.e.\ after 
the sub-pixel sampling, aspect ratio correction, and shift-and-add processes, 
all of which affect the noise distribution.
 Especially, the sub-pixel sampling has a significant effect 
as it divides one pixel into four pixels, resulting in 
signal in a pixel and thus $\sigma$ of its distribution
to be $1/4$ of the original values.
 We should also note here that 
this $\sigma$ corresponds to the standard deviation 
and differs from an uncertainty of the sky level in one image, 
which should be estimated by the standard error.
 Taking into account this sub-pixel sampling effect\footnote{
The $\sigma$ of an image after sub-pixel sampling is  
$\sigma{\rm (pix)}=0.25\sigma{\rm (pix)}_{\rm org}$ and 
the number of pixels used for aperture photometry is 
$N{\rm (pix)}=4N{\rm (pix)}_{\rm org}$, while the subscript ``org'' denotes 
those of an image before sub-pixel sampling.
The uncertainty in the aperture photometry is 
$\sigma{\rm (aper)}=\sqrt{N{\rm (pix)}_{\rm org}}\times \sigma{\rm (pix)}_{\rm org}
=2\sqrt{N{\rm (pix)}}\times \sigma{\rm (pix)}$.
}, 
we estimate the $5\sigma$ sensitivity in aperture photometry.
 Values for the apertures used in \citet{Tana08} are listed in Table \ref{tab:5sig_aper} 
in units of mJy.
\begin{table*}
  \tbl{
  Typical $5\sigma$ sensitivity in aperture photometry from NEP observations [mJy]
  \label{tab:5sig_aper}}{
    \begin{tabular}{lccccccccc}
      \hline
      AOT & N2 & N3 & N4 & S7 & S9W & S11 & L15 & L18W & L24 \\
      \hline
      IRC03 & 0.058 & 0.042 & 0.041 & 0.29 & 0.31 & 0.49 & 0.89 & 0.84 & 1.8\\
      IRC05 & 0.032 & 0.023 & 0.022 & 0.17 & 0.24 & 0.33 & 0.53 & 0.49 & 1.2\\
      \hline
    \end{tabular}}
\end{table*}

 We compare this sensitivity for IRC05 with that of WISE, whose 
four filters at 3.4, 4.6, 12, and 22 $\mu$m 
can be compared to N3, N4, S11, and L24, respectively.
 While WISE sensitivities depend on a position on the sky, 
the $5\sigma$ values for the COSMOS field \citep{Sco07} in the AllWISE Data 
Release\footnote{http://wise2.ipac.caltech.edu/docs/release/allwise/}
are 0.054, 0.071, 0.73, and 5.0 mJy, respectively.
 These values are $\sim 2$--4 larger than those for IRC05.

\subsection{PSF}\label{sec:PSF}
 The shape and size of the PSF in a stacked image 
depends on the instrumental PSF, the pointing stability, 
and the accuracy of relative shift calculations.
 While PSFs differ from the gaussian profile especially for MIR-L images \citep{Ari11}, 
we here provide information on PSF sizes by fitting a two-dimensional (2D) gaussian profile.

 We measure 
PSF sizes on stacked images from NEP observations 
performed during 
2006 October and 2006 December.
 Histograms of measured sizes (major and minor axis lengths 
in FWHM 
of the fitted 2D gaussian) are created 
for each filter and each AOT (see 
Appendix \ref{sec:psf}).
 Typical PSF sizes are estimated from the peak of these histograms
and listed in Table \ref{tab:PSF}.
 From this table, it is clear that NIR PSFs are more elongated 
than those of MIR.
 We attribute these NIR elongated PSFs to 
a larger amount of the jitter and/or drift during
longer exposure times 
compared to MIR.
 As summarized in Table \ref{tab:aot}, the
exposure time of one NIR long exposure frame for IRC03 and IRC05 
is 44.4 sec and 65.5 sec, respectively,
while that for one MIR long exposure frame 
is 16.4 sec for all AOTs 
(see also \cite{Tana08}).
 On the other hand, the number of frames used for stacking 
is typically a factor of two or more smaller for IRC03 than for IRC05. 
 If errors in shift calculations have a significant effect, 
PSF sizes of IRC05 should be larger than those of IRC03. 
 Although the NIR PSF sizes of IRC05 are slightly larger than those of IRC03, 
the difference is only comparable to the width of their histograms
and no such trend is seen in the PSF sizes for MIR (Table \ref{tab:PSF}).
 We thus conclude that 
the pointing uncertainty 
due to the jitter and/or drift 
during an exposure has a dominant effect on 
the PSF
shape rather than the uncertainty of relative shift calculations.
 In addition, we find that major axes of PSFs are roughly along 
the y-axis for NIR images and x-axis for MIR images. 
 These PSF elongation directions both correspond to 
the cross-scan direction of the satellite (see Figure \ref{fig:FoVs}).
 This result indicates that the drift was more significant in that direction.
 (See Appendix \ref{sec:psf} for more detail.)

\begin{table*}
  \tbl{Typical PSF size of stacked images in FWHM [arcsec] from NEP observations
  \label{tab:PSF}}{
    \begin{tabular}{lccccccccc}
      \hline
      AOT & N2 & N3 & N4 & S7 & S9W & S11 & L15 & L18W & L24\\
      \hline
      IRC03 & $3.9\times 5.4$ & $3.3\times 6.2$ & $3.3\times 6.0$ & 
      $6.3\times 5.0$ & $6.5\times 5.3$ & $6.4\times 5.5$ & 
      $6.2\times 5.4$ & $6.7\times 6.1$ & $7.5\times 6.5$\\
      IRC05 & $3.9\times 5.7$ & $3.4\times 6.4$ & $3.3\times 6.4$ &
      $6.2\times 4.9$ & $6.6\times 5.2$ & $6.8\times 5.5$ &
      $6.2\times 5.5$ & $6.7\times 5.8$ & $7.2\times 6.7$\\
      \hline
    \end{tabular}}
\end{table*}

\subsection{Confusion}
 Using the typical PSF sizes measured above and 
NEP observations performed with IRC05 and during 2006 August and 2007 April, 
we examine if stacked images are limited by the source confusion.
 We should note here that the source confusion condition (i.e.\ the local source density) 
depends primarily on the Galactic latitude.
 Since the NEP field is not close to the Galactic plane ($b\sim 30^\circ$), 
the confusion limit can be higher for observations toward lower Galactic latitudes.

 For a stacked image, we identify sources brighter than $5\sigma$ and perform aperture photometry.
 The aperture radius for a source and sky is the same as \citet{Tana08}.
 Following the procedures adopted by \citet{WadaT08}, 
the source number density per beam is calculated, where the beam area 
is given by $\pi ({\rm PSF_{maj}~[FWHM]}/2.35)\times ({\rm PSF_{min}~[FWHM]}/2.35)$ 
and ${\rm PSF_{maj}}$ and ${\rm PSF_{min}}$ are major and minor axis length 
of PSF listed in Table \ref{tab:PSF}, respectively.
 If this density exceeds $1/30$, 
we regard the image is confusion limited and derive the confusion limit flux
(c.f.\ \cite{Con74}; \cite{Hogg01}). 

 Among 161 pointed observations toward the NEP region,
$\sim 50$--60 stacked images are available for each filter.
 We find that none of the images are confusion 
limited except for one N2 and one N3 images that have severe artificial noises.
 Meanwhile, \citet{WadaT08} create a large mosaic of the NEP field 
combining $\sim 200$ {\it AKARI}/IRC pointed observations.
 For NIR, they estimate the $5\sigma$ sensitivity to be $\sim 10~\mu$Jy 
and find that the source number densities are close to the limit (i.e.\ $1/30$).
 Since our sensitivities for aperture photometry at NIR are about twice their values, 
our conclusion above is consistent with their results.

\subsection{Flux calibration stability}
 Following \citet{Tana08}, we use photometric standard stars 
to check the flux calibration factors 
and their stability during Phase 1\&2.
 The flux calibration factor is defined as the ratio of 
the flux predicted by models \citep{Coh96,Coh99,Coh03a,Coh03b,Coh03z} 
to the observed flux, which is measured by aperture photometry on 
frames after the aspect ratio correction.
 We find that the calibration factor was stable within $\sim$10\%, 
noting that the error includes uncertainties of the
standard star models.
 No significant dependence on time is found, as already reported by \citet{Tana08}.

\section{Remaining artifacts and issues}\label{sec:issues}
 In this section, we describe artifacts and issues 
still remaining in the processed images.
 All of them are also explained in more detail in the IRC 
data users manual.

\subsection{NIR detector anomalies}
 NIR detector anomalies such as column pulldown and muxbleed are not removed. 
 Both of them appear when a bright source falls into the detector FoV.
 The former is a decrease of the signal of pixels in the same column with the source, 
resulting in a negative vertical stripe.
 The latter is 
a cyclic pattern in the signal of pixels in the same row as 
the source, 
resulting in short stripes aligned horizontally.
 Note that these artifacts cannot be removed by dithering as they move with the source.
 \citet{Murata13} masked out columns and rows with an object 
brighter than a certain limit before stacking,
which worked well for their NEP data sets since the field was observed many times 
with different rotation angles.
 During one pointed observation, a rotation angle does not vary significantly
(Figure \ref{fig:hshift}),
so that these anomalies still remain in stacked images.

\subsection{Ghost patterns}
 The ghost is an artifact due to 
reflections between optical elements.
 As described in \S \ref{sec:MIRflat}, the extended ghost pattern for 
the background sky in MIR-L is removed during the flat calibration.
 Other ghost patterns due to bright compact objects still remain 
in the processed images.
 \citet{Ari11} identified small-scale ghosts that appear close ($\sim 1'$) to a bright object 
in MIR frames.
 It is also known that similar ghosts appear in NIR frames \citep{Murata13}.
 In addition to these close ghosts, a large-scale arc-like ghost pattern is found 
at $\sim 1^\circ$ from a bright object.
 The shape of this pattern appears to be different in different channels, 
but has not fully been investigated yet.

\subsection{Memory effect}
\begin{figure}
 \begin{center}
  \includegraphics[trim=140 40 270 0,clip,width=0.45\linewidth]{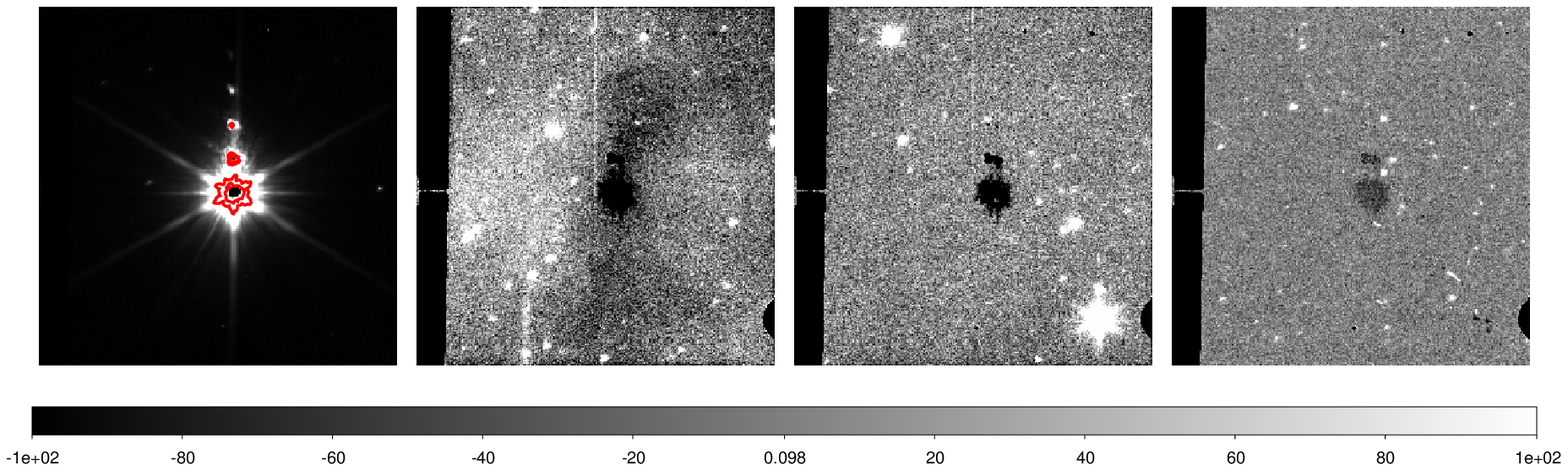} 
  \includegraphics[trim=140 29 270 0,clip,width=0.45\linewidth]{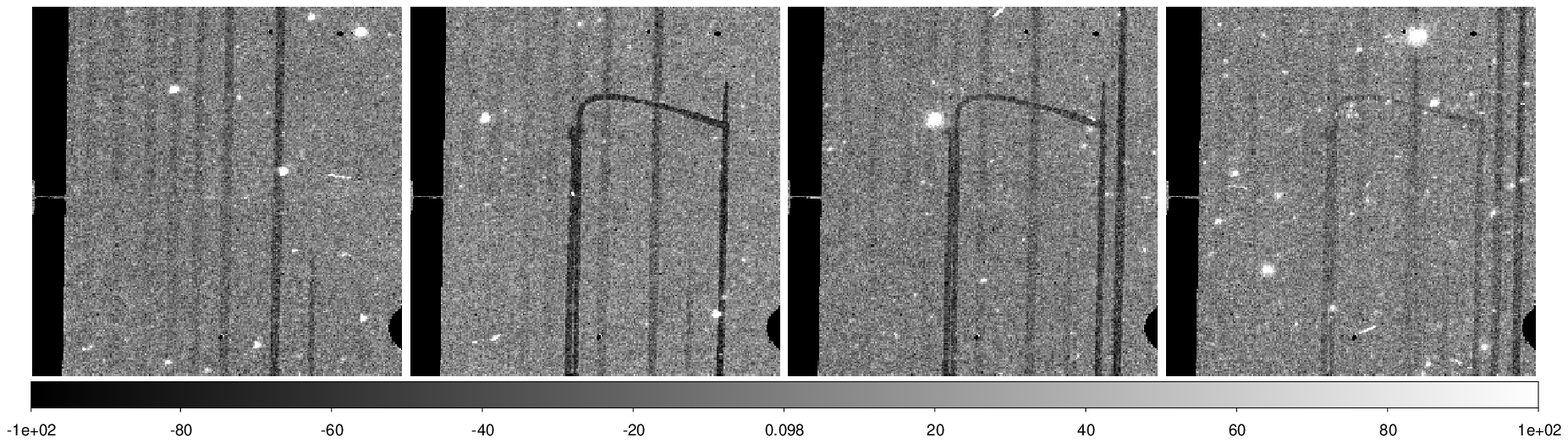} 
 \end{center}
\caption{S11 frames after flat correction as an example of memory effects. 
Left: a negative (dark) area at the center due to a bright star in a previous pointed observation.
Right: negative stripes due to one in the all-sky survey 
together with an arc-like pattern due to one in an FIS slow-scan observation.}
\label{fig:memory}
\end{figure}

 The memory effect is a temporal decrease of the pixel sensitivity 
after observing a bright object.
 This effect appears as dark spots, is most evident in MIR-S and sometimes in MIR-L, 
and lasts for several hours.
 The shape represents how the bright object was observed 
as presented in Figure \ref{fig:memory}.
 When a bright source was observed in a pointed observation, 
a negative area resembling the shape of the source appears in the following observations 
(Figure \ref{fig:memory}, left).
 Vertical stripes are seen (Figure \ref{fig:memory}, right) 
when a bright source was observed in the all-sky survey, 
as the telescope scanned the sky along the y-axis of MIR frames 
and the source was observed in several scans.
 An arc-like pattern and a partial stripe (Figure \ref{fig:memory}, right) represent how a bright source 
was observed in the slow-scan mode (i.e.\ during the parallel observations).
 The duration of memory effect is likely dependent on the source brightness and observing mode, 
but the number of data is not enough to establish the relationship.
 The amount of decrease is also not characterized, so that the affected pixels 
should be masked out.
 In some cases, saturation masks for the bright object are useful 
to mask the affected pixels out.
 Tasks for copying and applying these masks are available in the toolkit.

\subsection{Earthshine light}\label{sec:EL}
 When the angle between the satellite and the earth became smaller, 
stray light from the earth  
appeared in object frames.
 This is called earthshine light (EL)  and is most notable 
when observing an object around the NEP during summer.
 The problem is that it is extended but not uniform
and that its strength depends on the angular separation of the satellite and the earth limb.
 When the background sky level changes significantly during one pointed observation, 
it is most likely due to the EL.
 Assuming that the EL pattern shape does not change during a pointed observation, 
\citet{Egu13} created a template and successfully removed it from 
images with a nearby galaxy, i.e.\ extended object.
 Tasks for creating and subtracting an EL template are available in the toolkit.

\subsection{Drift}\label{sec:drift}
 The satellite pointing was controlled by two optical telescopes 
called star trackers installed on the satellite wall.
 In some observations large drifting took place during an exposure.
 Although the cause of such large drift is not fully understood yet, 
we deduce that two possibilities are
too few reference stars in the star tracker FoVs and
thermal distortion between the telescope axis and the reference frame of the attitude sensors.
 The PSFs of NIR long exposure frames become very elongated in these cases, 
since the exposure time is longer than that of MIR.
 In some extreme cases, the toolkit fails to identify sources used for stacking.
 While MIR PSFs are not so elongated compared to NIR,
the shift values of some frames deviate from the nominal values, 
and thus such frames are excluded from stacking (\S \ref{sec:calcshift}).
 As already mentioned,
this drift mostly occurred in the cross-scan direction.

\subsection{Stacking for multiple pointed observations}
 The current toolkit does not support stacking frames of
multiple pointed observations.
 A user can use the toolkit to produce processed frames before stacking 
for each pointed observation 
and then use his/her own code to stack all the processed frames from 
multiple observations.
 Note that the central coordinates and rotation angle may differ 
between observations 
with the same targetID, i.e.\ even when the same object was observed with the same AOT.

\subsection{Saturation}
 Bright objects become saturated in long-exposure frames.
 Such saturated pixels are masked out during the pipeline processing 
but may be recovered by using the 
short-exposure frames. 
 However, such a technique has not been established 
nor implemented in the toolkit.

\section{Summary}
 {\it AKARI} is the Japanese infrared astronomical satellite, which performed 
an all-sky survey and pointed observations with the IRC and the FIS.
 The IRC is equipped with nine photometric filters covering 2--27 $\mu$m 
continuously. 
 In this paper, we describe the latest calibration processes for IRC images 
from pointed observations in Phase 1\&2, when the telescope was 
cooled with liquid Helium.
 Especially, dark frames, flat frames, calculation schemes of relative shift values, 
and WCS matching have been improved.

 We use the latest toolkit for data reduction to process IRC images 
from $\sim 4000$ pointed observations during Phase 1\&2.
 Target objects include a wide range of sources from asteroids to distant galaxies.
 Through the pipeline processing, one stacked image for each filter 
used in one pointed observation is created.
 About 90\% of the stacked images have a position accuracy better than $1.5''$, 
while this percentage generally decreases with wavelength.
 Typical sensitivities are
estimated to be a factor of 
$\sim 2$--4 better than AllWISE at similar wavelengths.
 Remaining artifacts and issues in the processed images are also described.

 All of the products including the latest toolkit, processed images, 
process logs, and Data Users Manual are available via 
the {\it AKARI} Observers website.

\begin{ack}
 The authors fully appreciate a referee's careful reading of the manuscript 
and a number of helpful comments and suggestions to improve it.

This work is based on observations with {\it AKARI}, 
a JAXA project with the participation of ESA. 
 The authors greatly appreciate Dr.\ T.~Nakamura for his work on the WCS matching code.
 The authors also thank Dr.\ T.~Koga, Mr.\ K.~Sano, Mr.\ S.~Koyama, Mr.\ S.~Baba, and Mr. Y.~Matsuki 
for checking the WCS coordinates of stacked images visually.
 Dr.\ S.~Takita kindly provided the web interface for searching the released data sets.

 This research has made use of the VizieR catalogue access tool, CDS, Strasbourg, France.
 This publication makes use of data products from the Two Micron All Sky Survey, 
which is a joint project of the University of Massachusetts and 
the Infrared Processing and Analysis Center/California Institute of Technology, 
funded by the National Aeronautics and Space Administration and the National Science Foundation, 
and also from the Wide-field Infrared Survey Explorer, 
which is a joint project of the University of California, Los Angeles, 
and the Jet Propulsion Laboratory/California Institute of Technology, 
funded by the National Aeronautics and Space Administration.
\end{ack}

\appendix
\section{Details of released data sets}\label{sec:datasets}
 The processed data together with documents and the 
latest
toolkit was released on 
2015 March 31, 
with a minor update 
to data on 2015 April 30.
 All of the products are available via the {\it AKARI} observers website.
 In Table \ref{tab:obslist}, we list a proposal title, 5-digit proposal code, proposal type, 
and the number of ObsIDs included in this release.
 
 The data package for one ObsID contains neighbor dark frames, 
a Readme, a log of pipeline processing, and stacked images.
 The full options adopted during the process are recorded in the log.
 The Readme file provides a summary for the observation 
and for the pipeline processing such as the number of frames used for stacking 
and uncertainties in the WCS matching.

 A summary file for all the released ObsIDs is also available.
 This file lists a summary of information in the Readme for one ObsID per line.
 Users can overview the whole observations in Phase 1\&2 and 
select ObsIDs that satisfy certain criteria.
 In addition to a web interface for searching the data sets by object names and coordinates,
we provide a list of observations based on the on-sky positions
on proposals.
 The list of proposals is also available online.

\begin{table*}
  \caption{Title, Proposal ID, type, and the number of ObsIDs for proposal in Phase 1\&2}
  \label{tab:obslist}
\small   
  \begin{center}
    \begin{tabular}{lllc}
      \hline
      Title & Code & Type$^\dagger$ & \# of ObsIDs \\
      \hline
     North Ecliptic Pole Survey & LSNEP & LS & 718\\
      Large Magellanic Cloud Survey & LSLMC & LS & 598\\
 Evolution of Cluster of Galaxies & CLEVL & MP & 278 \\
 Interstellar dust and gas in various environments of our Galaxy and nearby galaxies & ISMGN & MP & 260 \\
 ASTRO-F Studies on Star formation and Star forming regions & AFSAS & MP & 237 \\
 DT for IRC & DTIRC & DT & 224 \\
 Debris Disks Around Main Sequence Stars and Extra-solar Zodiacal Emissions & VEGAD & MP & 197 \\
 Origin and Evolution of Solar System Objects & SOSOS & MP & 157 \\
 Unbiased Slit-Less Spectroscopic Survey of Galaxies & SPICY & MP & 140 \\
 Astro-F Ultra-Deep Imaging/Spectroscopy of the Spitzer/IRAC Dark Field & EGAMI & OT & 87 \\
 Mass loss and stellar evolution in the AGB phase & AGBGA & MP & 81 \\
 Evolution of ULIRGs and AGNs & AGNUL & MP & 65 \\
 Understanding the Dust Properties of Elliptical Galaxies & EGALS & OT & 52 \\
 FUHYU - SPITZER WELL STUDIED FIELD MISSION PROGRAM & FUHYU & MP & 50 \\
 Making AGN come of age as cosmological probes & 6AND7 & OT & 50 \\
 The Nature of New ULIRGs at intermediate redshift & NULIZ & OT & 48 \\
 NIR--MIR Spectroscopic Survey of Selected Areas in the Galactic Plane & SPECS & OT & 47 \\
 The spatial distribution of ices in Spitzer-selected molecular cores & IMAPE & OT & 46 \\
 PV for IRC & PVIRC & DT & 43 \\
 Astro-F Spectroscopic Observation of $z = 6$ QSOs & HZQSO & OT & 43 \\
 Search for Giant Planets around White Dwarfs & SGPWD & OT & 41 \\
 Deep IR Imaging of the Unique NEP Cluster at $z=0.813$ & CLNEP & OT & 38 \\
 15 Micron Imaging of Extended Groth Strip & GROTH & OT & 31 \\
 Formation and Evolution of Interstellar Ice & ISICE & OT & 30 \\
 Dust, PAHs and molecules in molecular clouds & CERN2 & OT & 28 \\
 UV-selected Lyman Break Galaxies at $0.6<z<1.3$ in the Spitzer FLS & Z1LBG & OT & 27 \\
 Mid-Infrared Imaging of the ASTRO-F Deep SEP Field & IRSEP & OT & 26 \\
 Mapping the Spectral Energy Distributions of Sub-mm Bright QSOs & SUBMM & OT & 22 \\
 Extreme Colors: The smoking Gun of Dust Aggregation and Fragmentation & COLVN & OT & 21 \\
 An ultra-deep survey through a well-constrained lensing cluster & A2218 & OT & 19 \\
 IRC NIR Spectroscopy of High-Redshift Quasars & NSPHQ & OT & 16 \\
 Cool Dust in the Environments of Evolved Massive Stars & WRENV & OT & 15 \\
 Near Infrared Spectroscopy of L and T Dwarfs & NIRLT & MP & 15 \\
 DT for FIS & DTFIS & DT & 13 \\
 Triggered massive-star formation in the Galaxy & AZTSF & OT & 13 \\
 The role of pulsation in mass loss along the Asymptotic Giant Branch & SMCPM & OT & 12 \\
 Dust and gas properties in AGB and post-AGB objects & CERN1 & OT & 12 \\
 Evolution of dust and gas in Photodissociation Regions & DGPDR & OT & 11 \\
 Activity of Small Solar System Bodies far from the Sun & ADAMB & OT & 11 \\
 Deep Extinction Maps of Dense Cores & DEMDC & OT & 10 \\
 Stars departing from the Asymptotic Giant Branch & DEAGB & OT & 10 \\
 Probing molecular tori n Obscured AGN through CO Absorption & COABS & OT & 10 \\
 A Search for Very Low Luminosity Objects in Dense Molecular Cores & VELLO & OT & 9 \\
 Search for Emission Outside the Disks of Edge-on Galaxies & HALOS & OT & 9 \\
 Accretion and protoplanetary disks in brown dwarfs & DISKB & OT & 8 \\
 The hidden evolution from AGB stars to PNe as seen by ASTRO-F/IRC & AGBPN & OT & 8 \\
 ASTRO-F/IRC Slit-less Spectroscopy of Hickson Compact Groups & SHARP & OT & 7 \\
 Far-infrared Emission from the Coma Cluster of Galaxies & FIREC & OT & 7 \\
 Far-Infrared Spectroscopic Observation of Eta Carinae & ETASP & OT & 7 \\
 Excavating Mass Loss History in Extended Dust Shells of Evolved Stars & MLHES & MP & 5 \\
 Calibrating mid-IR dust attenuation tracers for LBGs with ASTRO-F & IRLBG & OT & 5 \\
 Dusty Star-Formation History of the Universe & GALEV & MP & 5 \\
 Spectroscopic Search for Atmosphere of an Extra-Solar Planet & EXOSP & OT & 4 \\
 The stellar mass and the obscured star formation harboured by EROs & EROMU & OT & 4 \\
 15 Micron Imaging of the $z=2.38$ Filament & BLOBS & OT & 4 \\
 FIR Spectroscopy of Super Nova Remnants & SNRBS & OT & 3 \\
 Star Formation and Environment in the COSMOS Field & COSMS & OT & 3 \\
 Near-Infrared Spectroscopy of the Atmospheres of Uranus and Neptune & UNIRC & OT & 2 \\
 The Nature of the Dusty Medium in Dwarf Elliptical Galaxies & DUDES & OT & 2 \\
      \hline
    \multicolumn{4}{l}{$\dagger$: LS=Large Survey, MP=Mission Program, OT=Open Time, 
   DT=Director's Time}\\
    \end{tabular}
  \end{center}
\end{table*}
\renewcommand{\arraystretch}{1}  

\section{Details of the all-data processing}\label{sec:moredetail}
\subsection{Relative shift between frames}\label{sec:stat_0203}
\begin{figure*}
 \begin{minipage}{0.67\linewidth}
\includegraphics[width=0.45\linewidth]{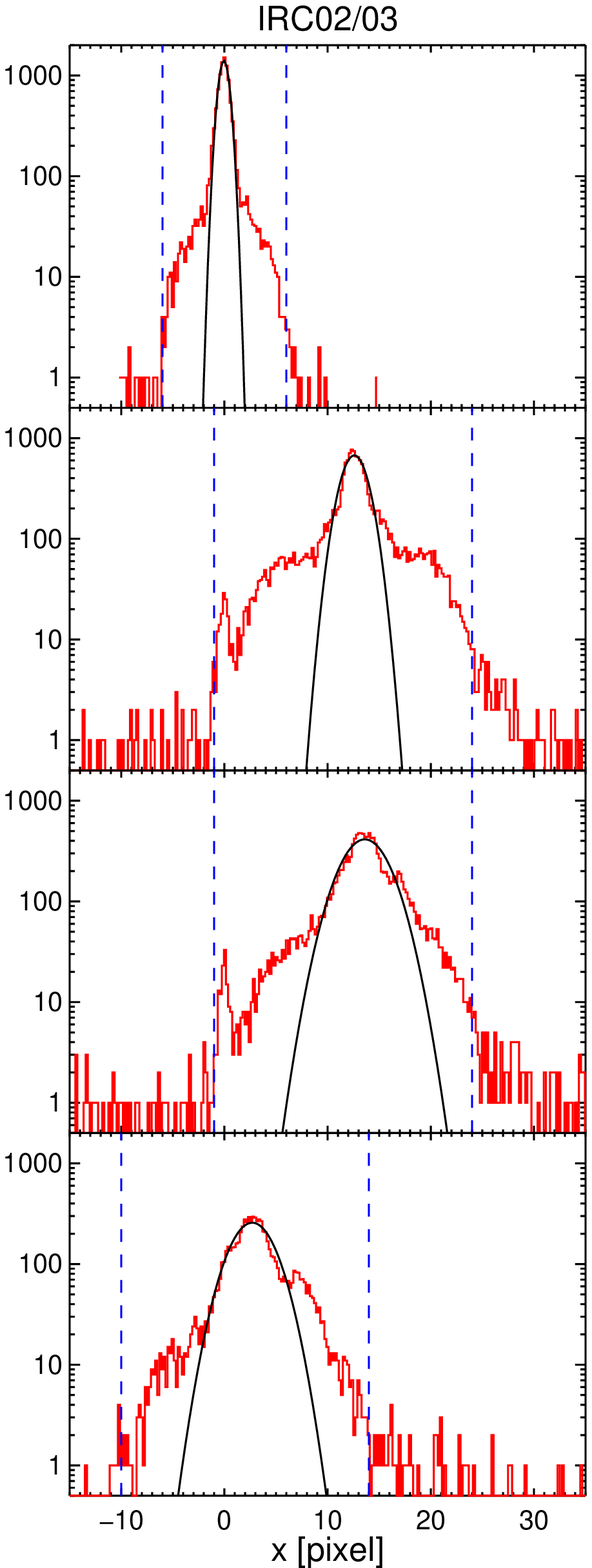} 
\includegraphics[width=0.45\linewidth]{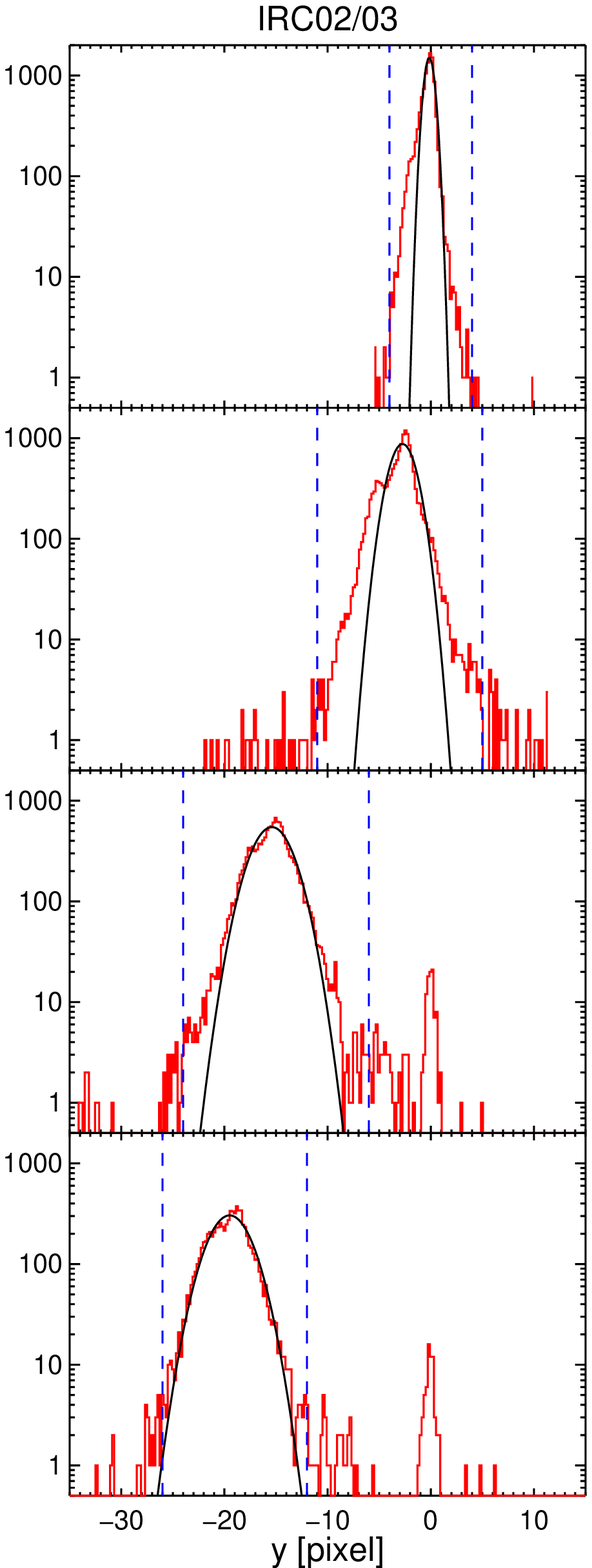} 
\caption{
Histogram of relative shift in x- (left) and y-direction (right) of MIR-S frames 
for each dithering cycle 
(from top to bottom) of IRC02 and IRC03.
Pixel 
numbers
are after the sub-pixel sampling, i.e.\ twice the original value.
One pixel is $1.17''$ for MIR-S.
Black curves represent the results of gaussian fit to the histogram 
and blue dashed lines indicate the lower and upper limits adopted in the toolkit.}
\label{fig:stat_xy0203}
\end{minipage}
\begin{minipage}{.3\linewidth}
\includegraphics[width=\linewidth]{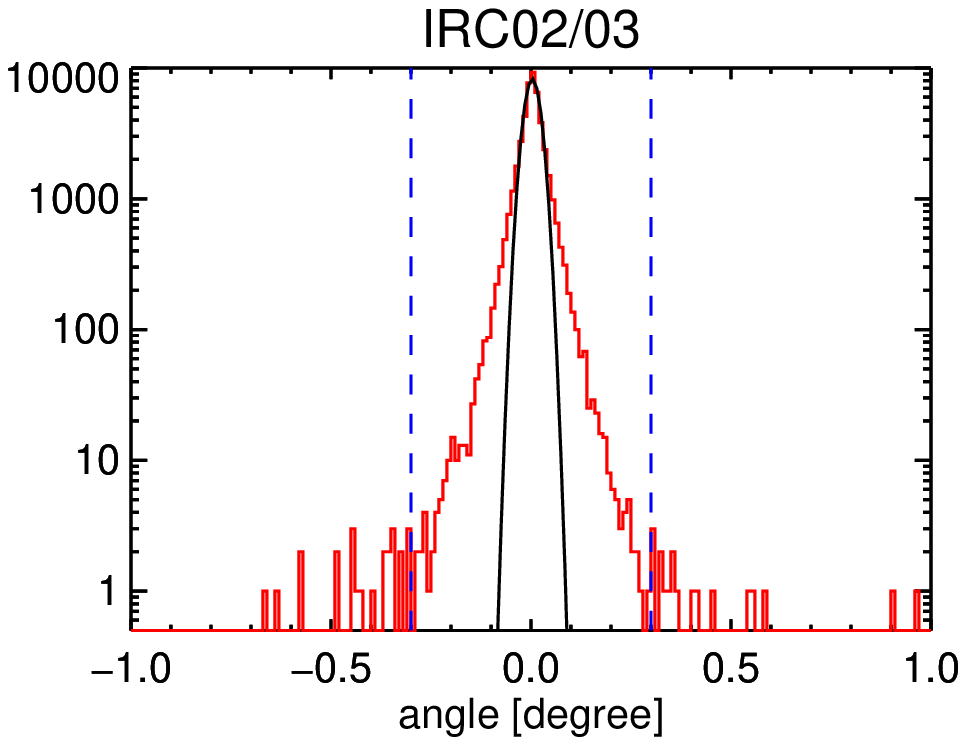} 
\caption{
Histogram of rotation angle of MIR-S frames for IRC02 and IRC03.
Black curves represent the results of gaussian fit to the histogram 
and blue dashed lines indicate the lower and upper limits adopted in the toolkit.}
\label{fig:stat_a0203}
\end{minipage}
\end{figure*}

 Histograms of relative shift in the x- and y-direction for each dithering cycle 
in the case of IRC02 and IRC03 are presented in Figure \ref{fig:stat_xy0203}.
 For the rotation angle, a histogram is created using the data from 
all dithering cycles (Figure \ref{fig:stat_a0203}).
 As with the case for IRC00 and IRC05 (Figure \ref{fig:hshift}), 
shift values are for MIR-S frames after the sub-pixel sampling and 
the lower and upper limits adopted in the toolkit are indicated by 
blue dashed lines.
 The wings in histograms of the x-direction shift 
found in the IRC00 and IRC05 cases are also seen in Figure \ref{fig:stat_xy0203}.

\subsection{Accuracy of WCS matching}\label{sec:wcserr}
\begin{figure*}
 \begin{center}
\includegraphics[trim=10 0 0 0,clip,width=0.32\linewidth]{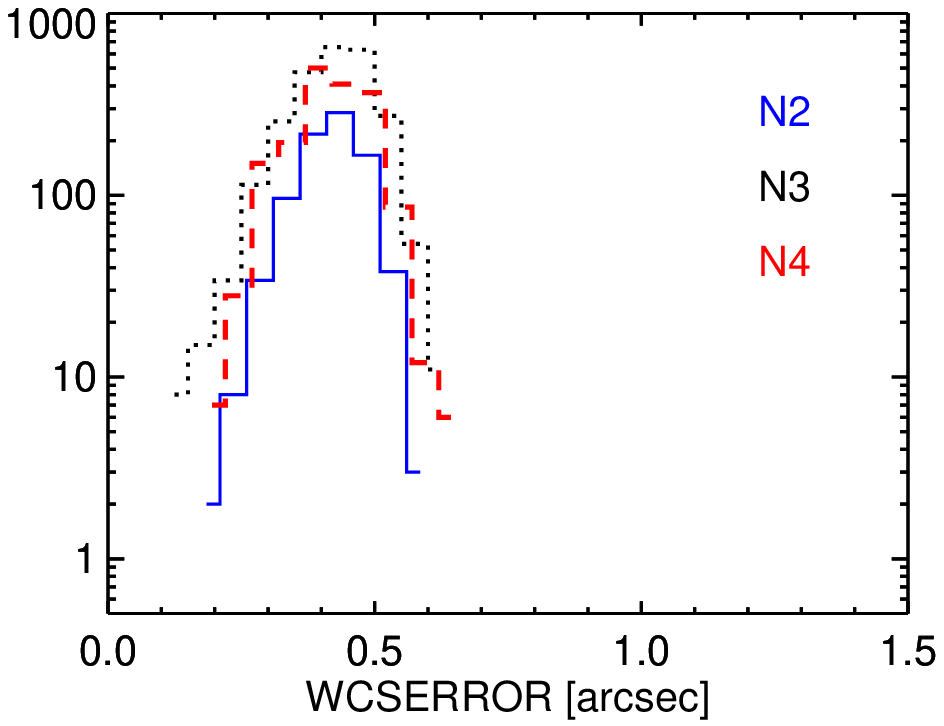} 
\includegraphics[trim=10 0 0 0,clip,width=0.32\linewidth]{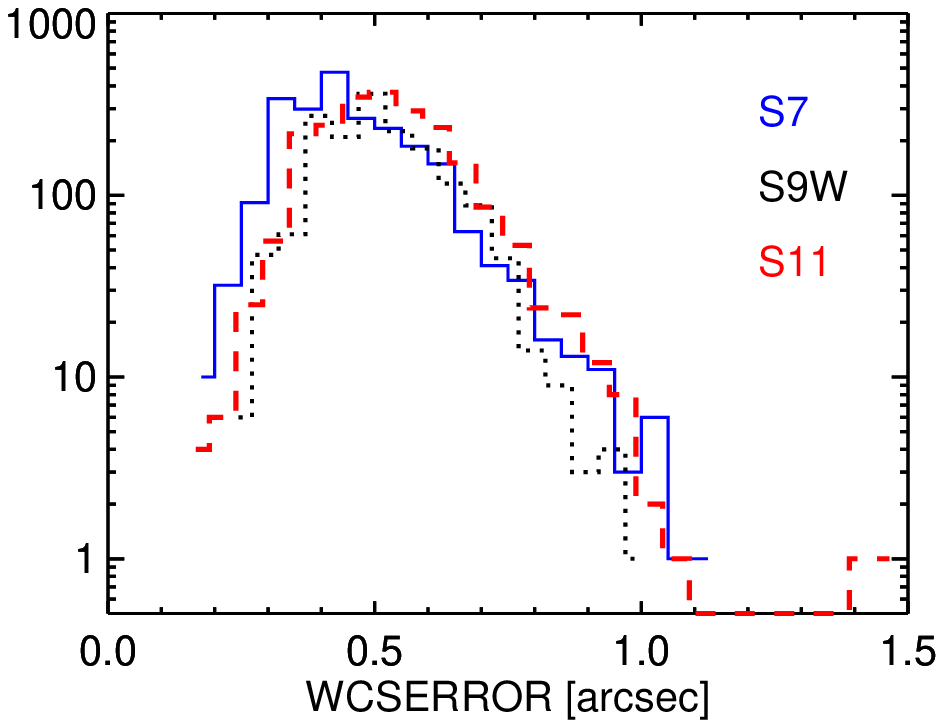} 
\includegraphics[trim=10 0 0 0,clip,width=0.32\linewidth]{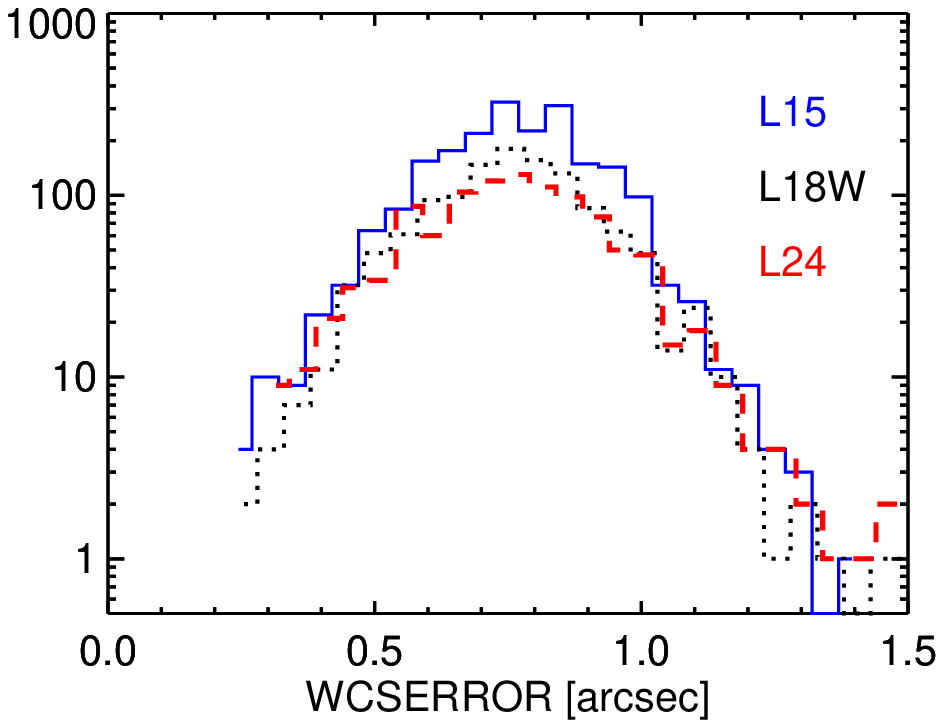} 
 \end{center}
\caption{
Histogram of residuals of the WCS matching for NIR (left), 
MIR-S (middle), and MIR-L (right).}
\label{fig:hwcserr}
\end{figure*}

 The residual of WCS matching is recorded as a FITS header parameter `WCSERROR'.
 This is a root-mean-square of positional offsets between 
matched WCS and catalogued WCS for sources in a stacked image.
 A histogram of this value for each filter is presented in Figure \ref{fig:hwcserr}.
 Since the tolerance for WCS matching is set to be $1.5''$, 
the residual is smaller than this limit.
 If the residual becomes larger than this limit, that WCS matching is considered as being failed. 

\subsection{Sensitivity variation}\label{sec:sigma_var}
\begin{figure*}
 \begin{center}
\includegraphics[trim=0 0 0 0,clip,width=0.32\linewidth]{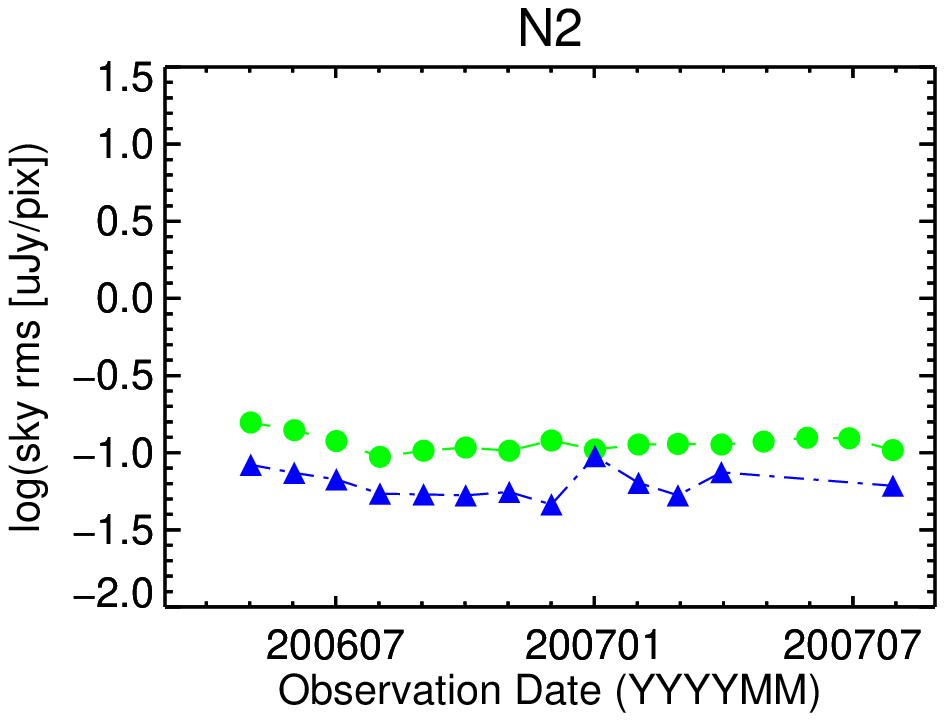} 
\includegraphics[trim=0 0 0 0,clip,width=0.32\linewidth]{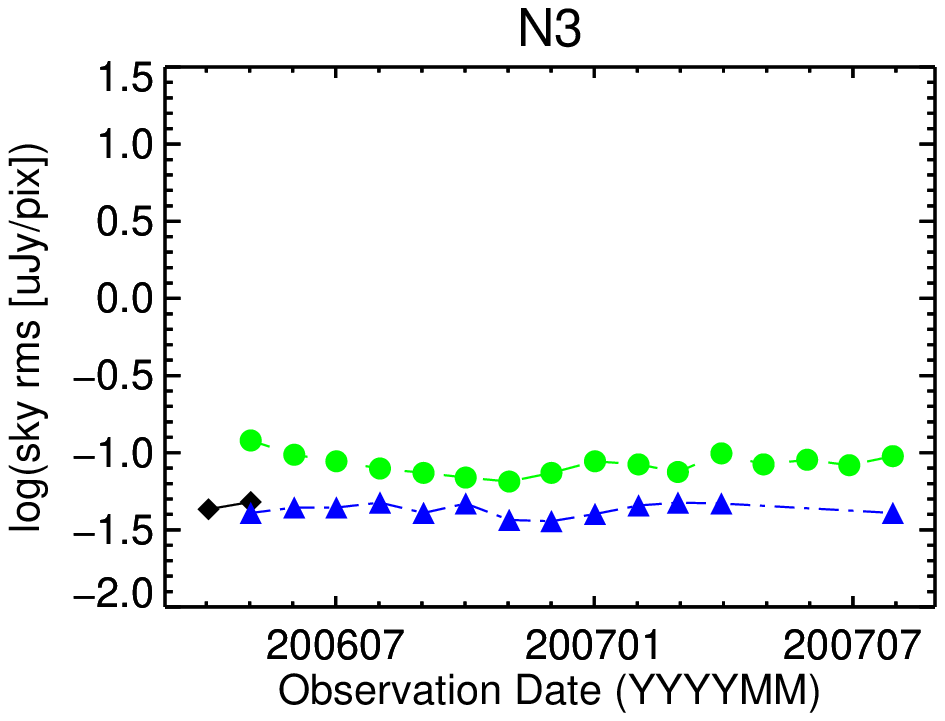} 
\includegraphics[trim=0 0 0 0,clip,width=0.32\linewidth]{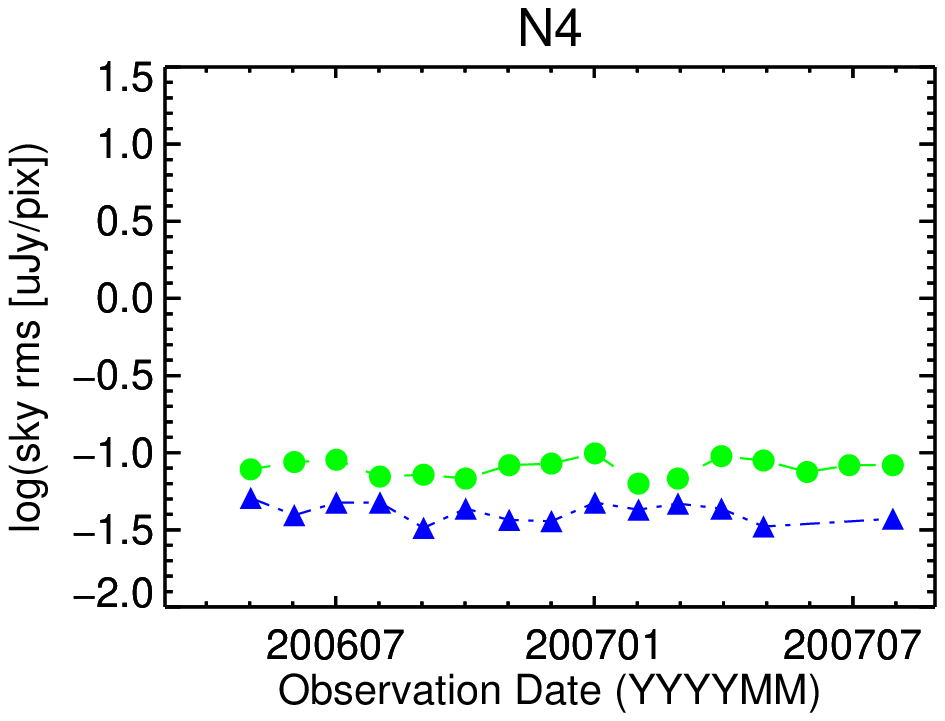} \\
\includegraphics[trim=0 0 0 0,clip,width=0.32\linewidth]{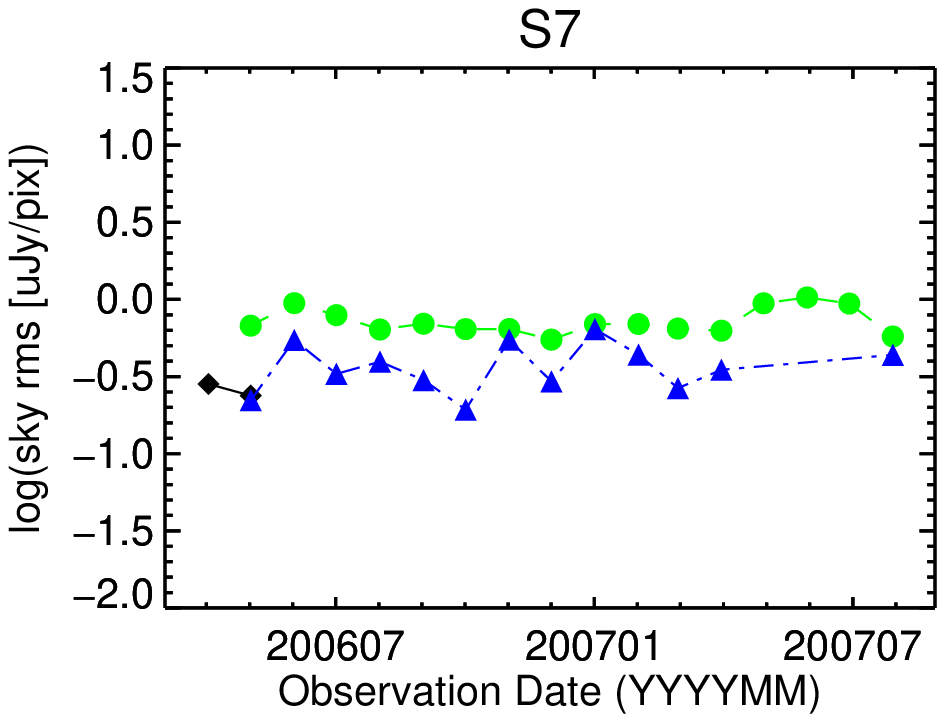} 
\includegraphics[trim=0 0 0 0,clip,width=0.32\linewidth]{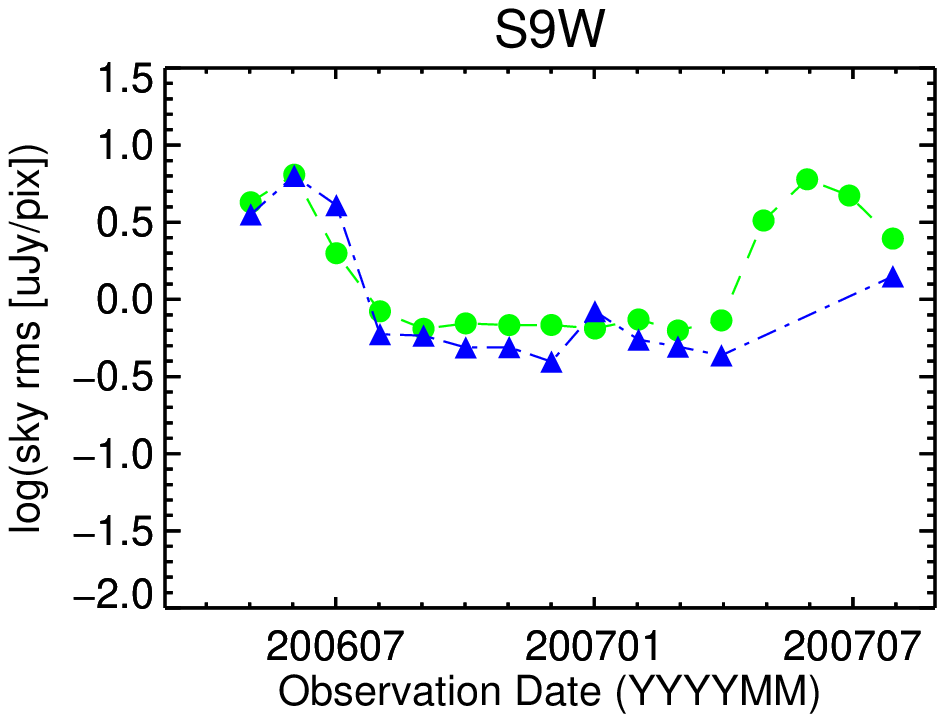} 
\includegraphics[trim=0 0 0 0,clip,width=0.32\linewidth]{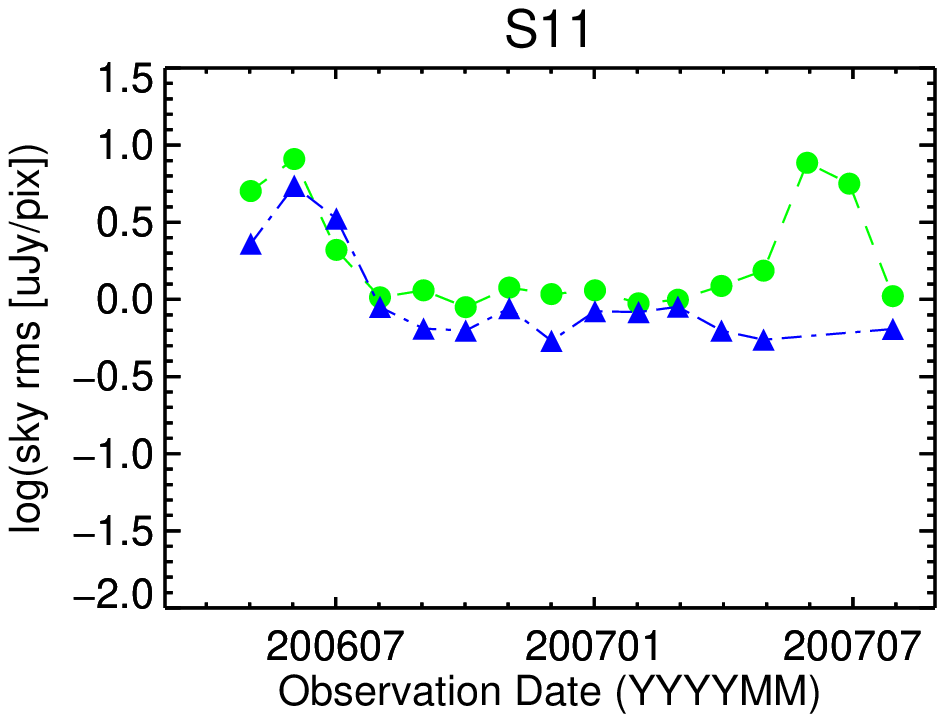} \\
\includegraphics[trim=0 0 0 0,clip,width=0.32\linewidth]{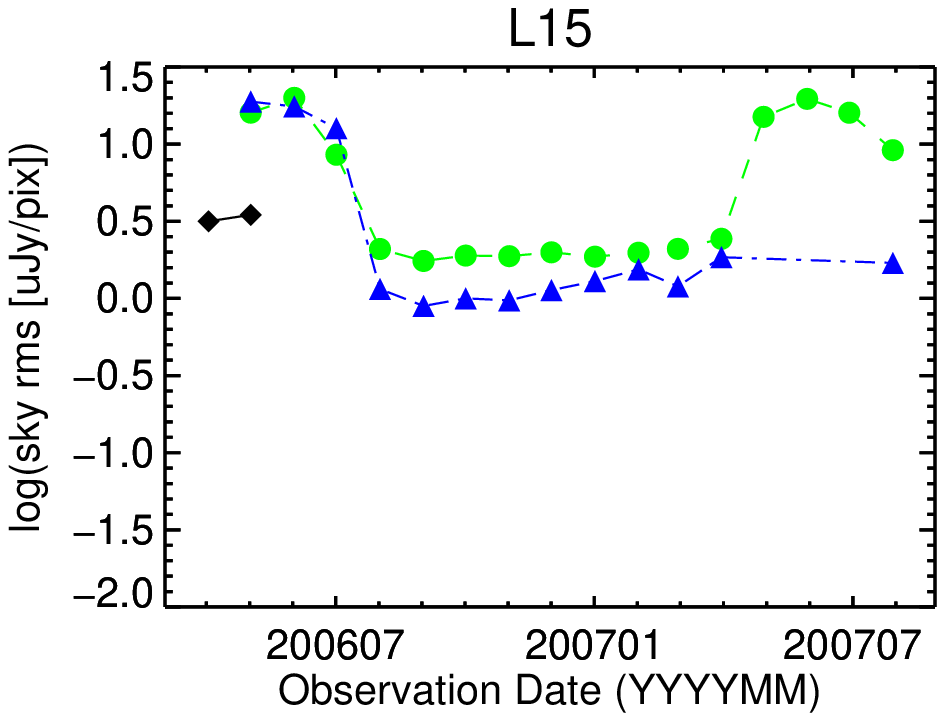} 
\includegraphics[trim=0 0 0 0,clip,width=0.32\linewidth]{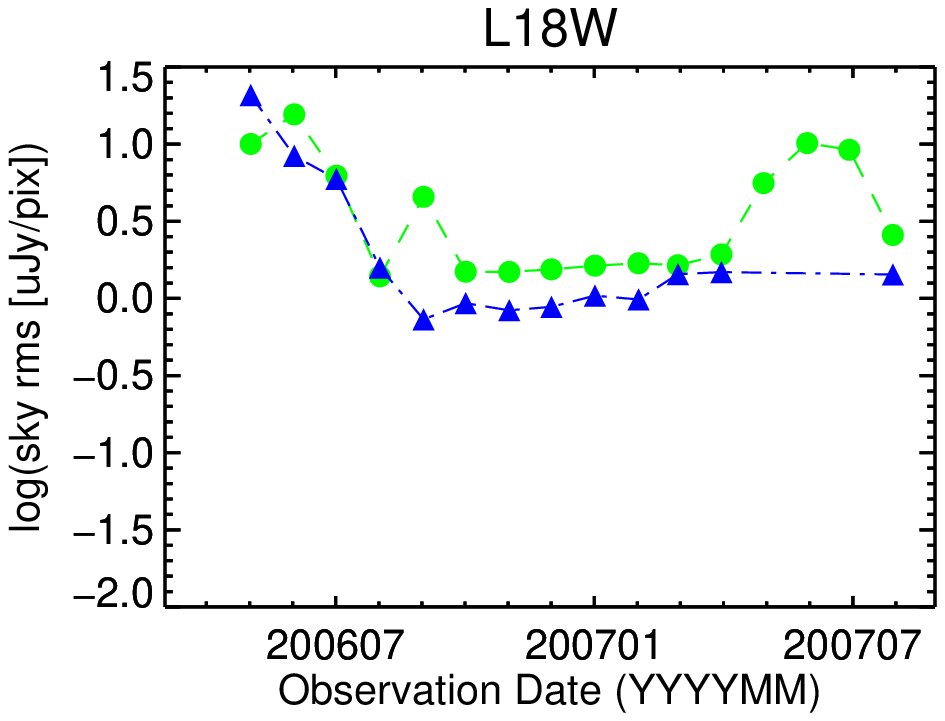} 
\includegraphics[trim=0 0 0 0,clip,width=0.32\linewidth]{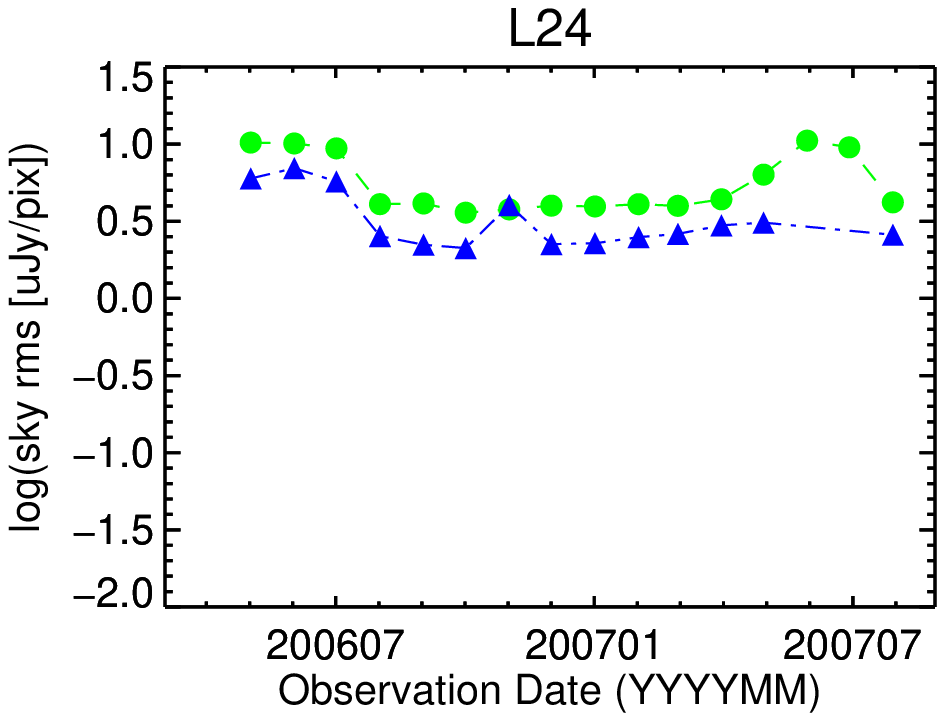} \\
 \end{center}
\caption{
Temporal variation of sky standard deviation from NEP observations.
Different symbols indicate different AOTs: black diamond for IRC00, green circle for IRC03, 
and blue triangle for IRC05.
}
\label{fig:sigma_NEP}
\end{figure*}

 As explained in \S \ref{sec:sen}, we measure the sky standard deviation $\sigma$
of stacked images from pointed observations toward the NEP region.
 For each month, AOT, and filter, we pick one pointed observation 
that has the largest number of stacked frames and measure $\sigma$.
 In Figure \ref{fig:sigma_NEP}, 
the temporal variation of $\sigma$ values is presented for the whole Phase 1\&2.
 We here note again that these values are measured in stacked images with sub-pixel sampling.
 Large $\sigma$ values during May and July at longer wavelengths are primarily due to 
the Earthshine light (\S \ref{sec:EL}).
 Therefore, we measure {\it typical} values of $\sigma$ (listed in Table \ref{tab:skyrms}) 
from data taken between 2006 August and 2007 April.

\subsection{Typical PSF sizes}\label{sec:psf}
\begin{figure*}
 \begin{center}
\includegraphics[trim=0 15 0 0,clip,width=0.3\linewidth]{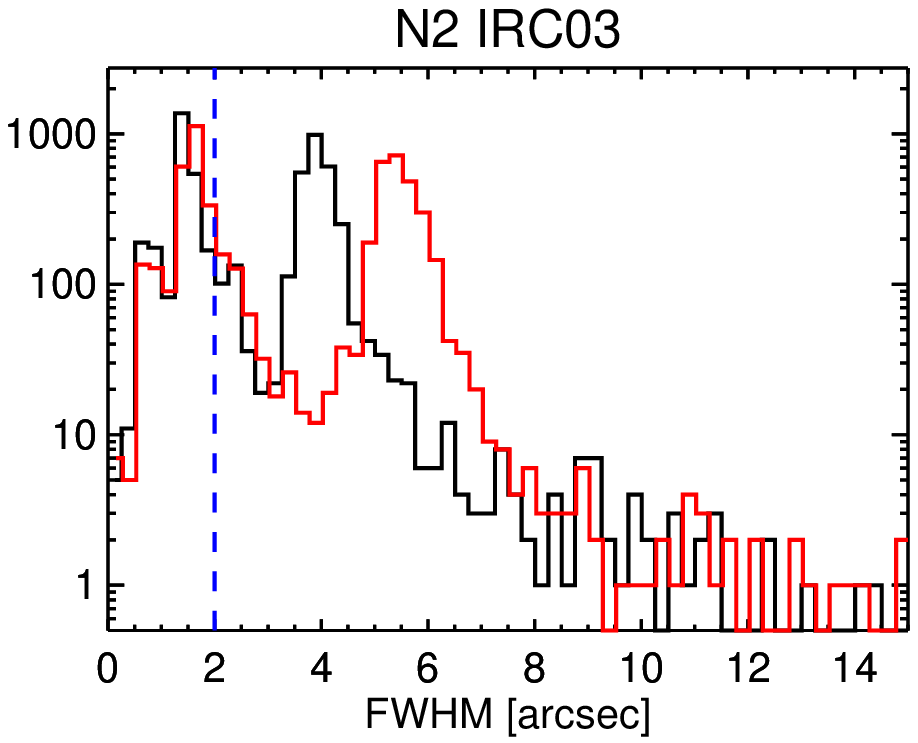} 
\includegraphics[trim=0 15 0 0,clip,width=0.3\linewidth]{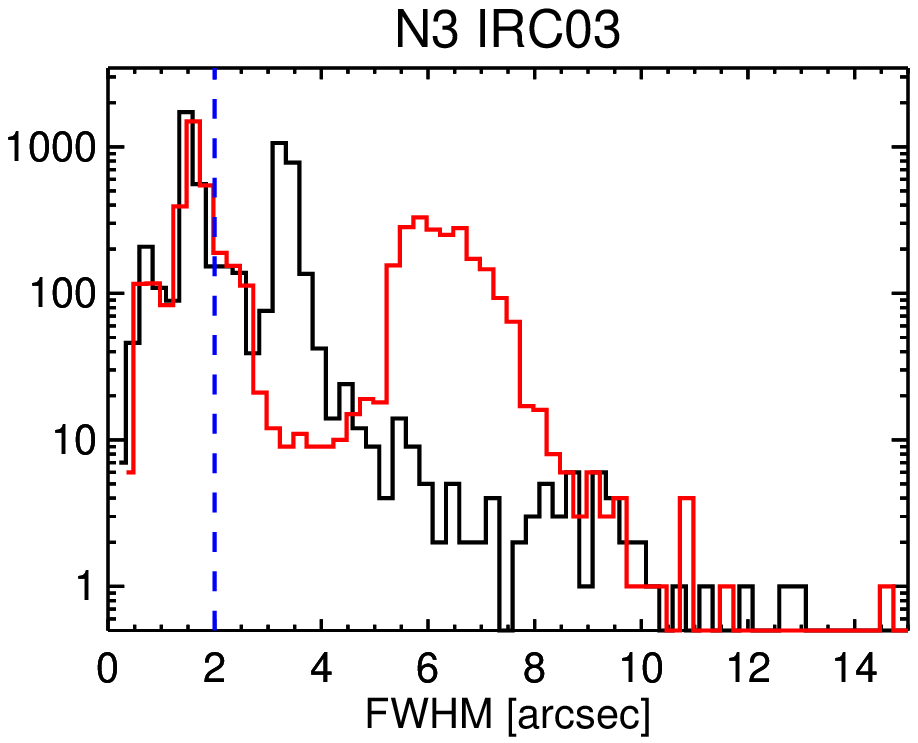} 
\includegraphics[trim=0 15 0 0,clip,width=0.3\linewidth]{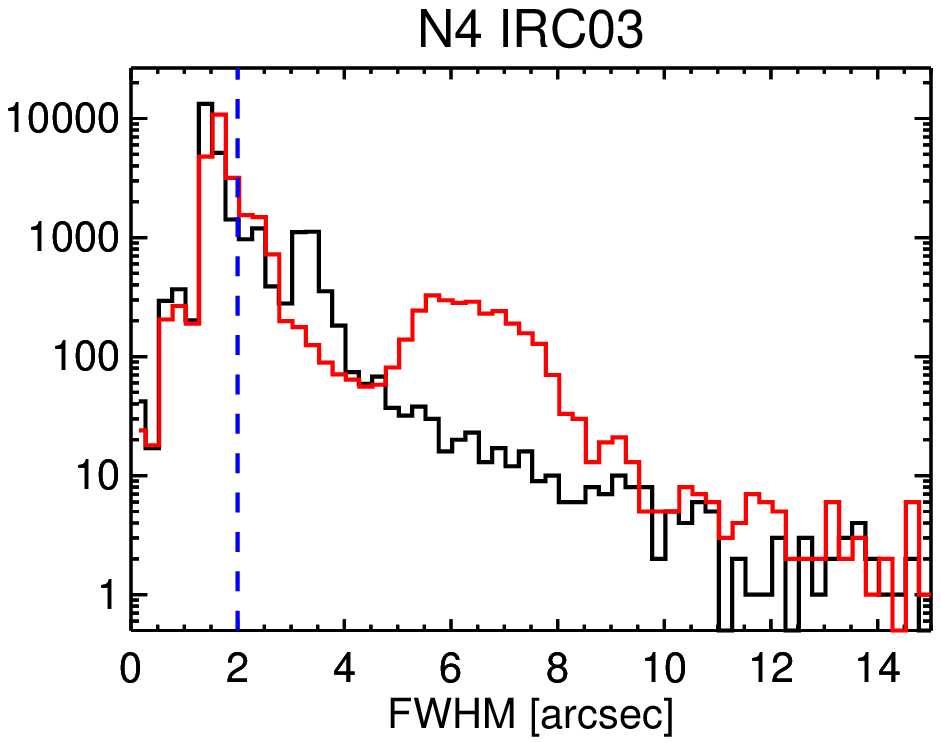} 
\includegraphics[width=0.3\linewidth]{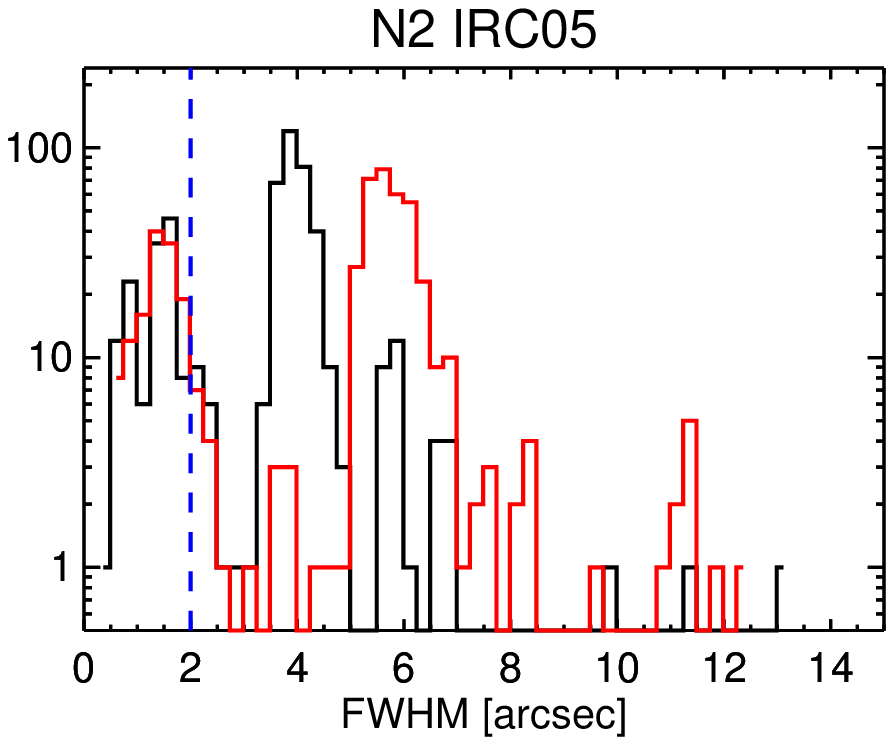} 
\includegraphics[width=0.3\linewidth]{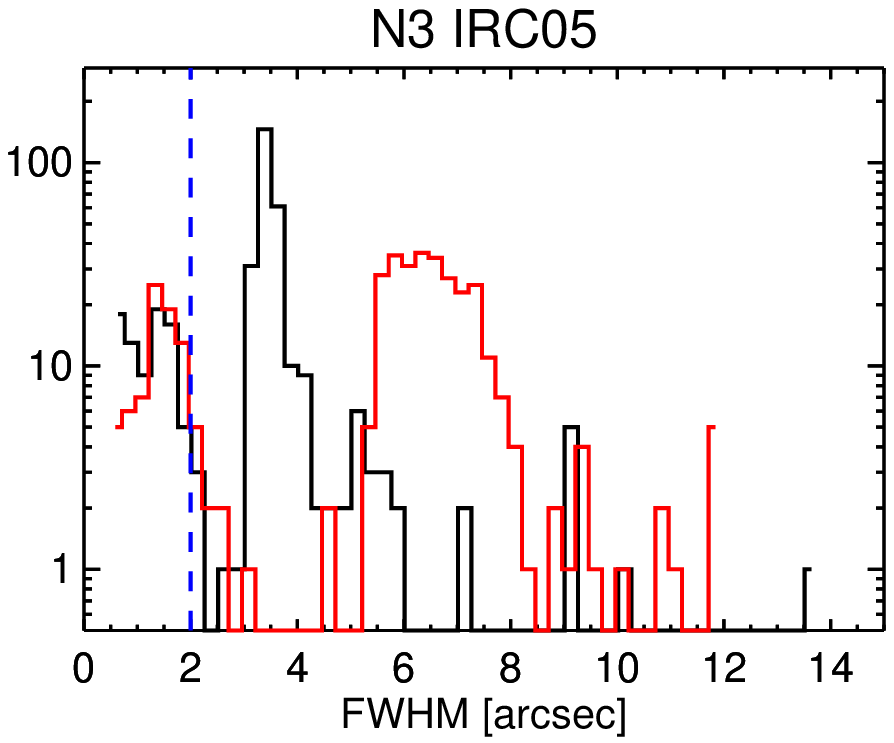} 
\includegraphics[width=0.3\linewidth]{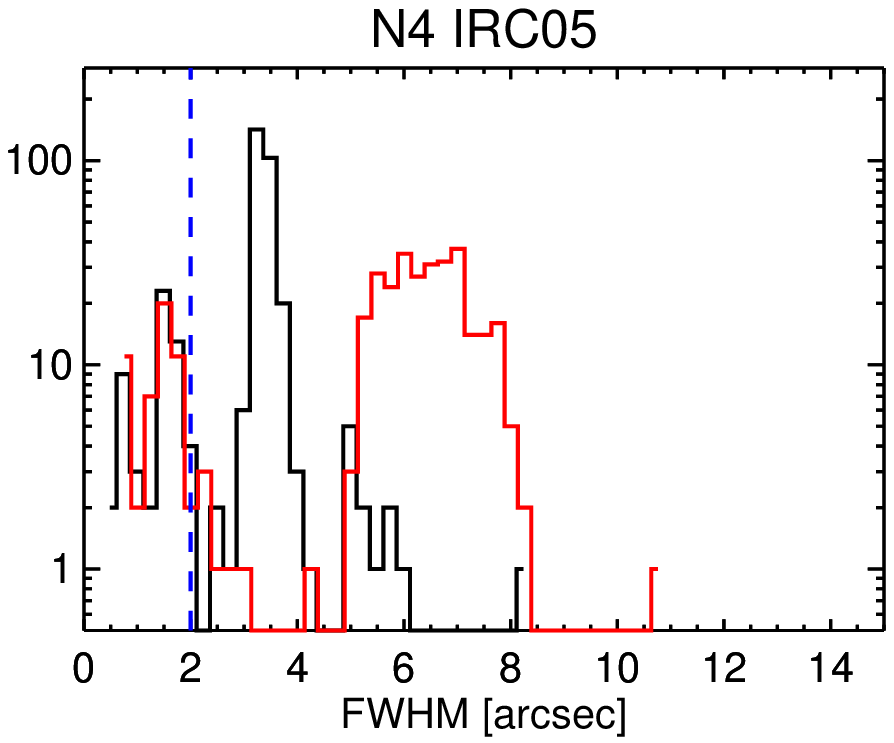} \\
\vspace{12pt}
\includegraphics[trim=0 15 0 0,clip,width=0.3\linewidth]{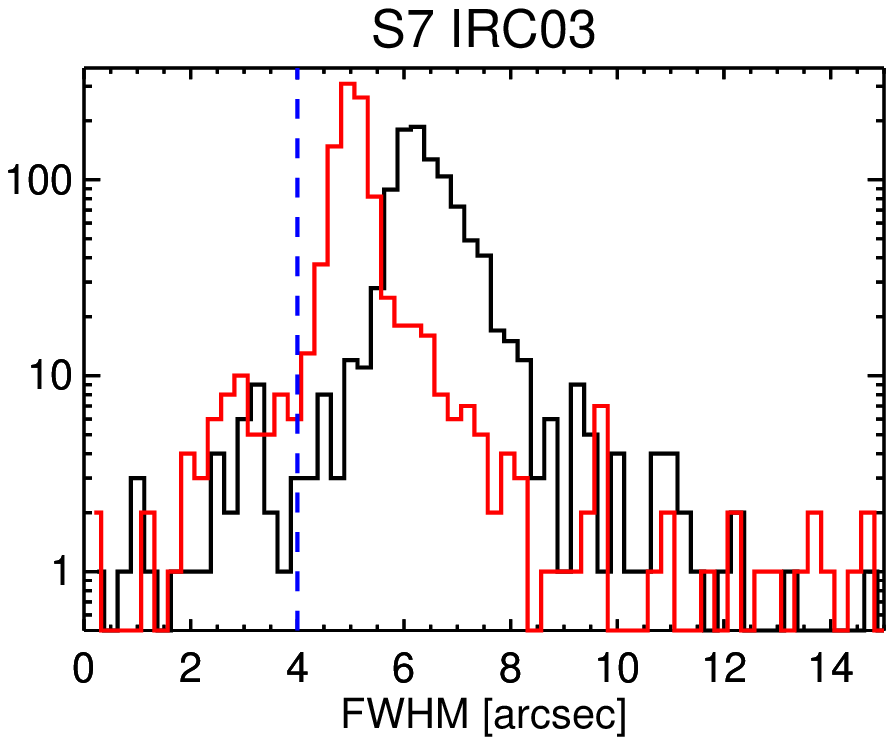} 
\includegraphics[trim=0 15 0 0,clip,width=0.3\linewidth]{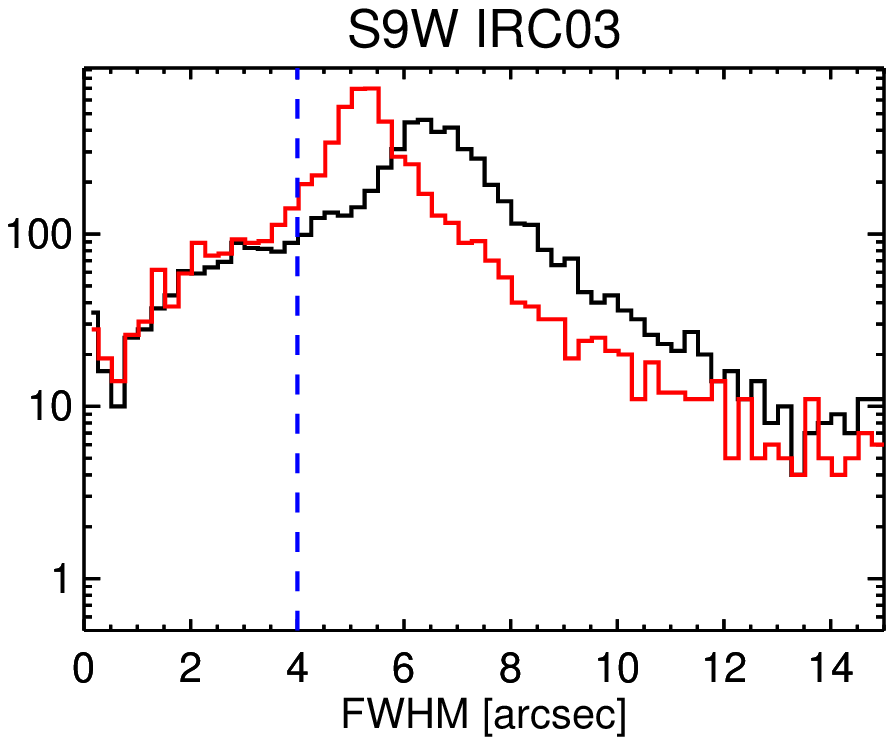} 
\includegraphics[trim=0 15 0 0,clip,width=0.3\linewidth]{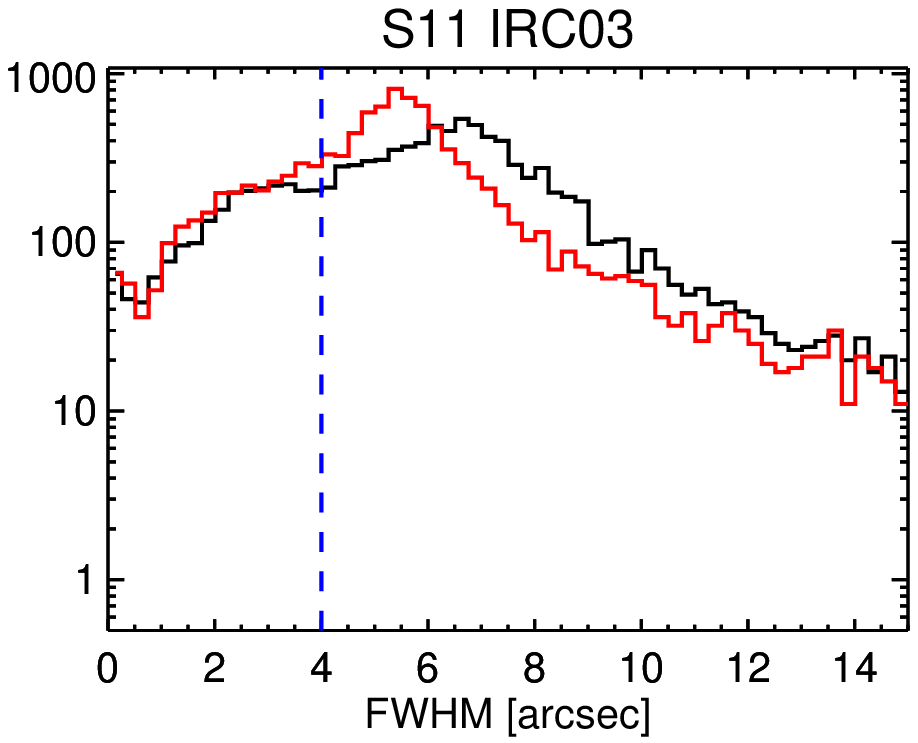} 
\includegraphics[width=0.3\linewidth]{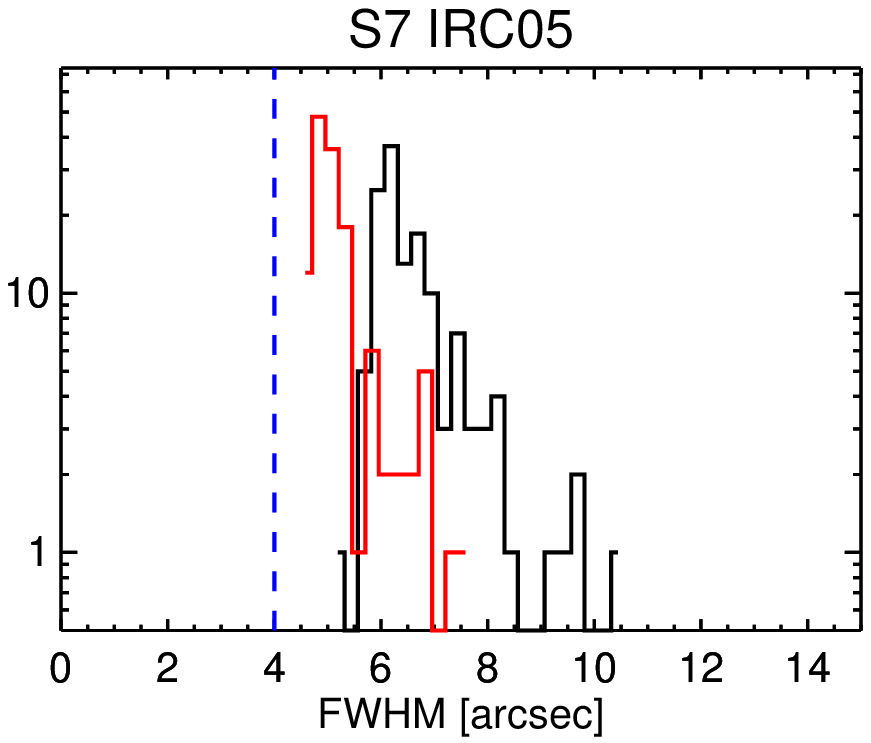} 
\includegraphics[width=0.3\linewidth]{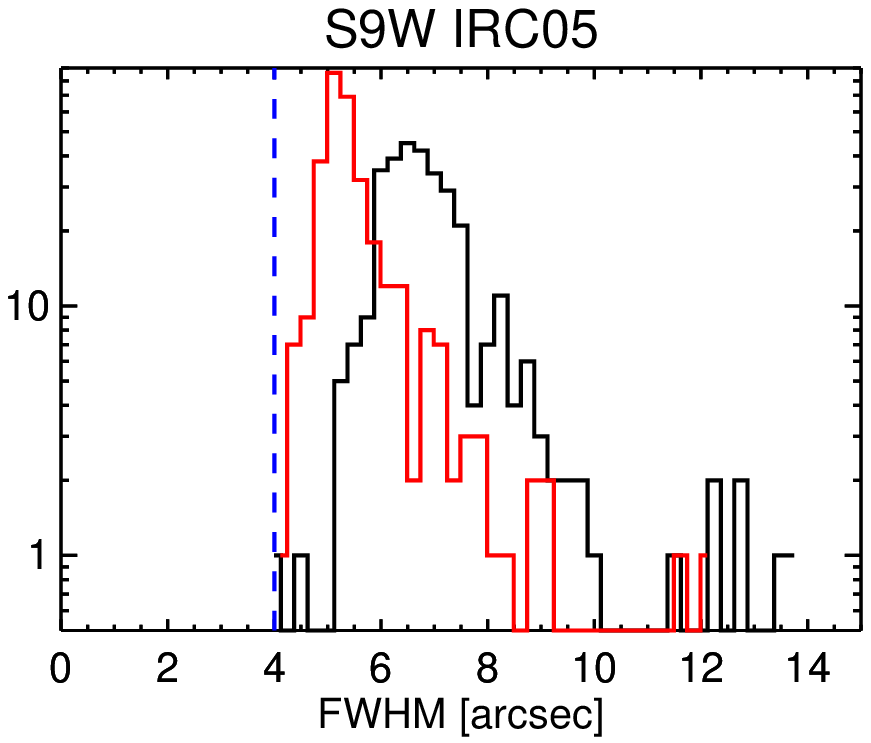} 
\includegraphics[width=0.3\linewidth]{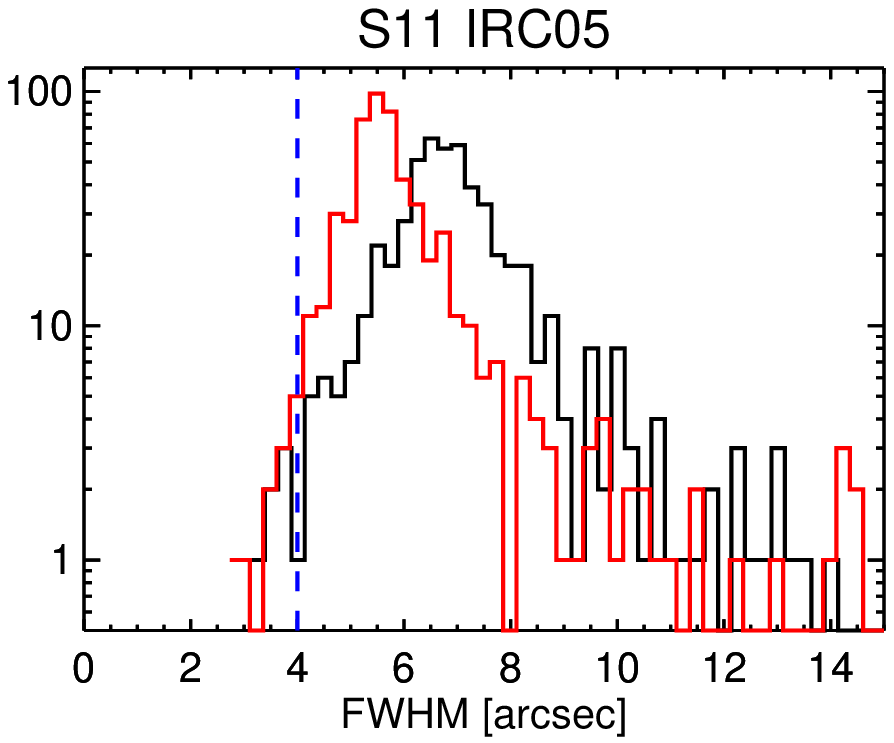} \\
\vspace{12pt}
\includegraphics[trim=0 15 0 0,clip,width=0.3\linewidth]{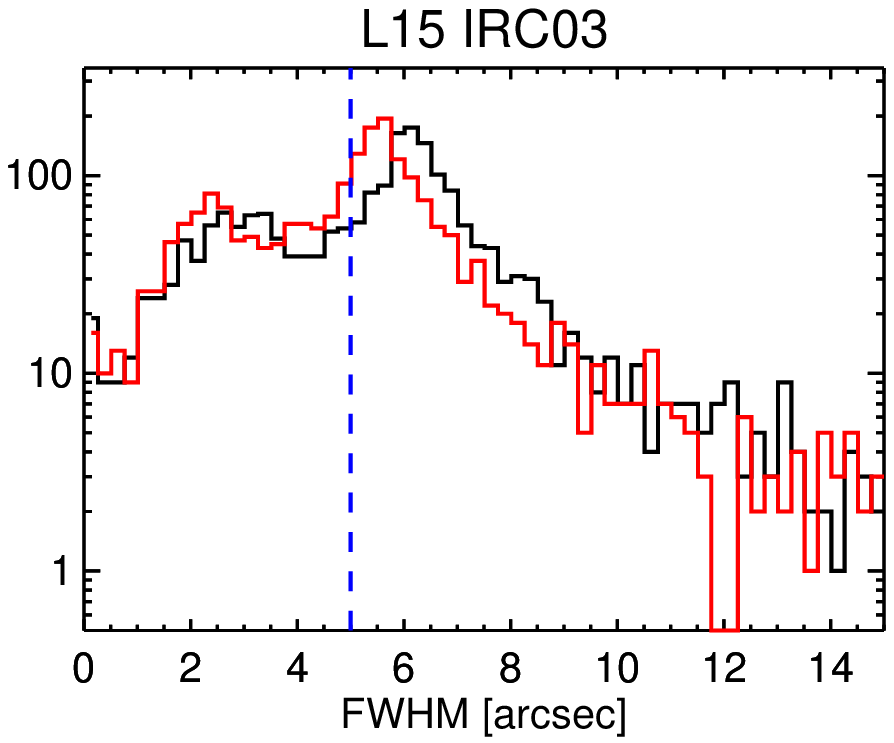} 
\includegraphics[trim=0 15 0 0,clip,width=0.3\linewidth]{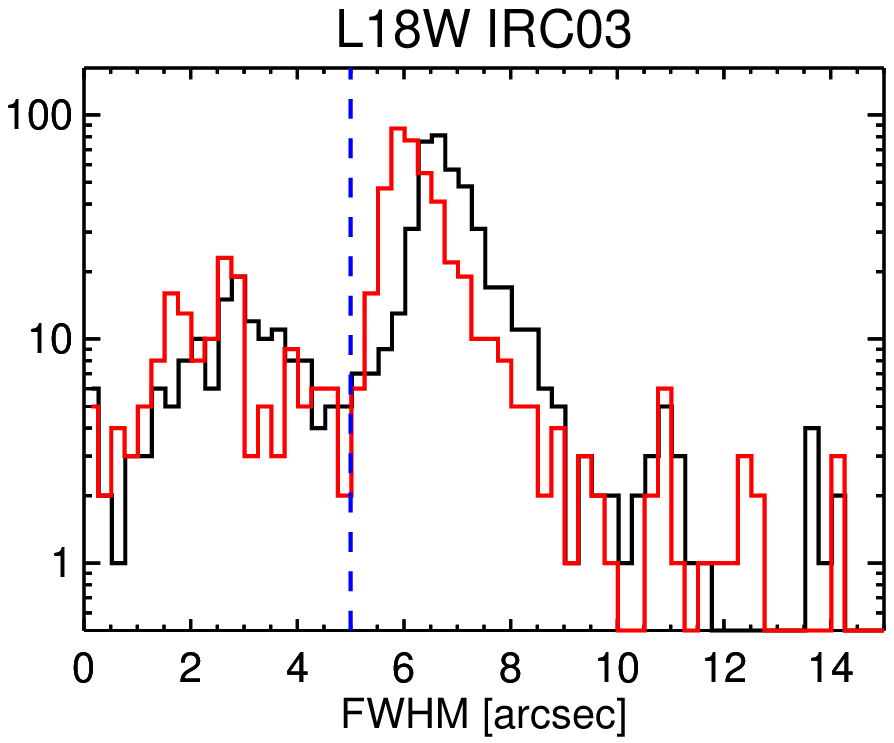} 
\includegraphics[trim=0 15 0 0,clip,width=0.3\linewidth]{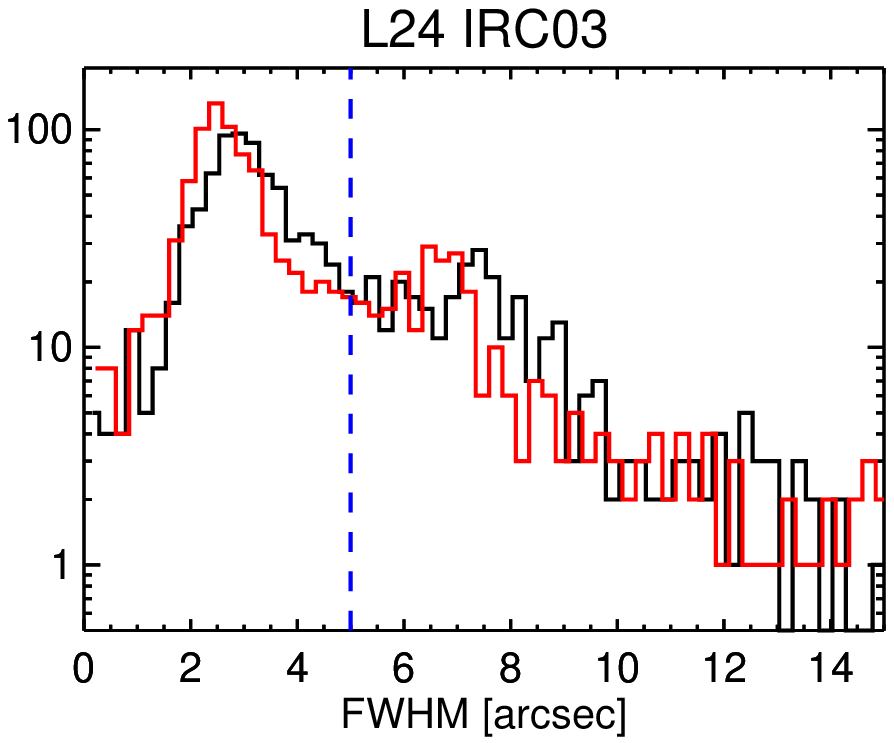} 
\includegraphics[width=0.3\linewidth]{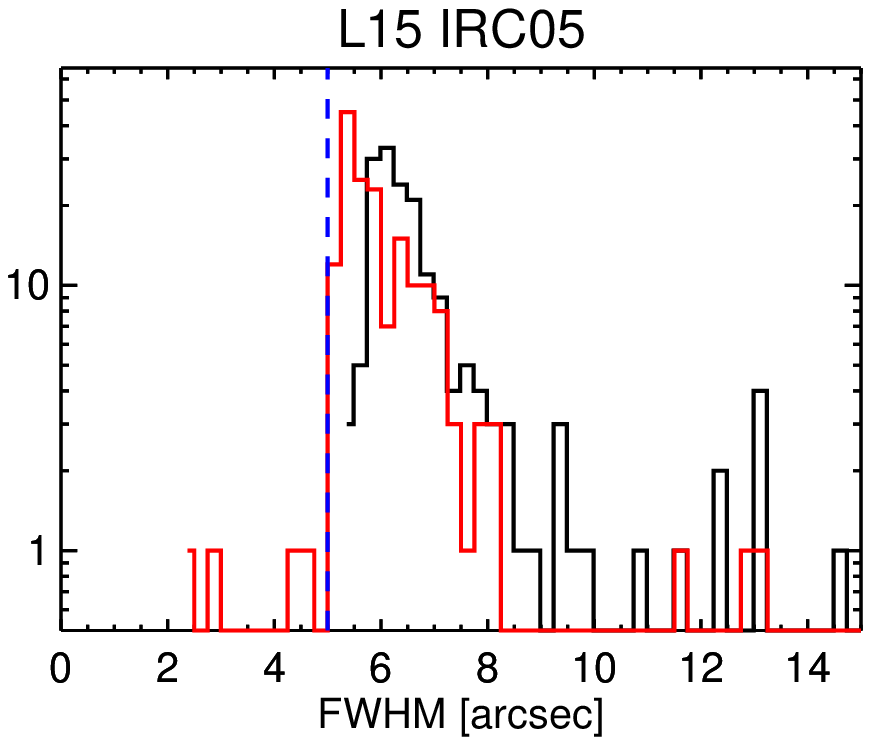} 
\includegraphics[width=0.3\linewidth]{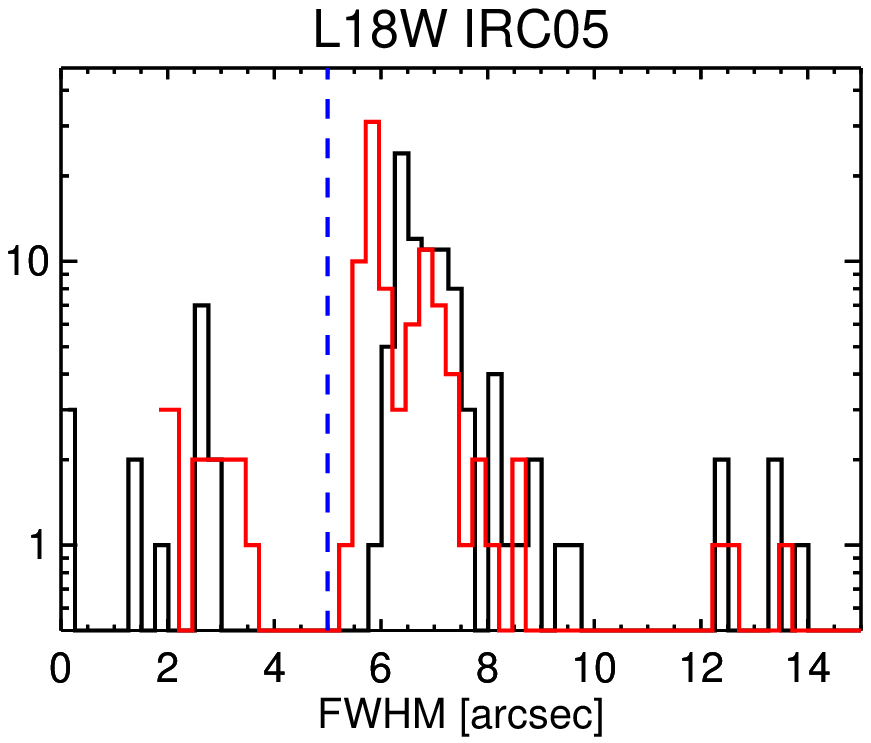} 
\includegraphics[width=0.3\linewidth]{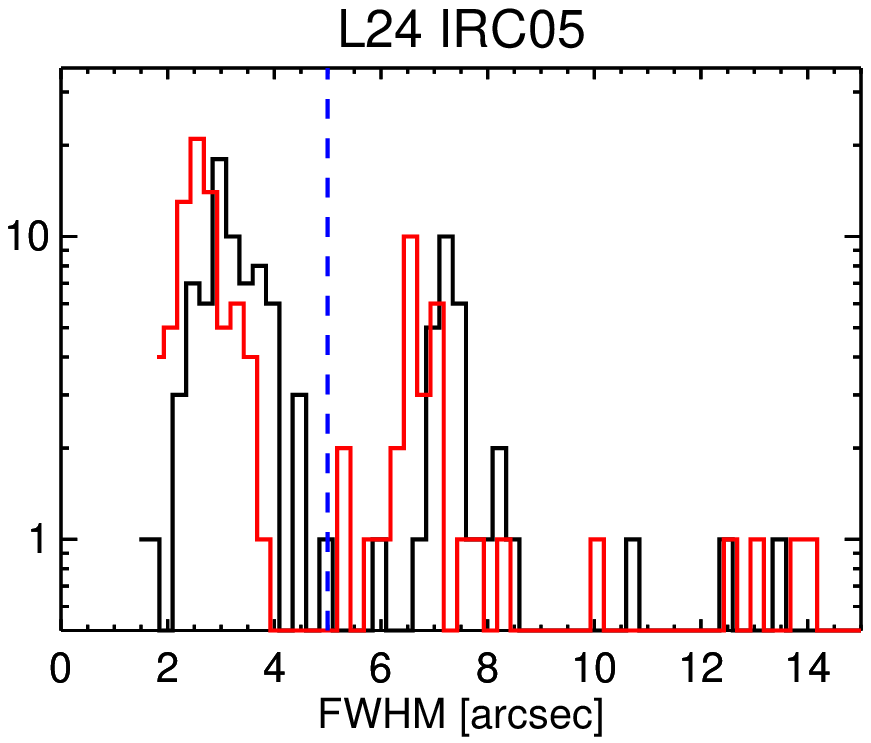} 
 \end{center}
\caption{
Histogram of PSF sizes of stacked images 
from NEP observations during 
2006 October--December.
FWHM [arcsec] along the major and minor axes of fitted 2D gaussian is presented.
Black and red lines indicate the length of axis closer to x- and y-directions 
of the stacked images, respectively.
Blue dashed lines indicate the lower limit adopted when fitting a gaussian to the histogram 
to estimate the typical PSF sizes listed in Table \ref{tab:PSF}.}
\label{fig:psf}
\end{figure*}

 In order to estimate PSF sizes of processed images, 
we performed a 2D gaussian fit to sources found in stacked images of NEP observations 
taken during 
2006 October and 2006 December.
 Histograms of measured sizes (major and minor axis lengths 
in FWHM
of the fitted 2D gaussian) were created 
for each filter and each AOT (Figure \ref{fig:psf}).
 Then a gaussian profile was fitted to each histogram and its peak position 
is adopted as a typical PSF size 
(Table \ref{tab:PSF}).
 During the fit, we manually set the lower limit of PSF sizes 
in order to exclude histogram peaks 
due to noises (especially in NIR of IRC03) and sub-peaks of PSFs (especially in L24).
 Widths of histograms and thus of fitted gaussian vary, 
but are typically $0.2''$--$0.5''$.

 In Figure \ref{fig:psf}, the black line corresponds to the length of axis 
closer to the x-axis of stacked images, while the red line corresponds to 
the one closer to the y-axis.
 For NIR, the peak of red histograms (i.e.\ the PSF size roughly along the y-axis) 
is always at larger values compared to that of black histograms.
 For MIR, the trend is reversed, i.e.\ that of black histograms is at larger values, 
although the difference is smaller than in the NIR cases.
 In summary, the NIR PSF is more elongated 
along the y-axis while the MIR PSF is slightly elongated along the x-axis.
 Considering the FoV alignment (Figure \ref{fig:FoVs}), 
both of these axes correspond to the cross-scan direction, 
which indicates that the effect of drift is more significant in this direction.
 This is consistent with the wings seen in histograms of relative shift values in 
the x-axis direction of MIR-S images 
(Figures \ref{fig:hshift} and \ref{fig:stat_xy0203}).

\bibliographystyle{apj}
\bibliography{papers}

\end{document}